\documentclass[11pt,a4paper]{article}

\usepackage[T1]{fontenc}
\usepackage[utf8]{inputenc}
\usepackage[margin=1in]{geometry}
\usepackage{amsmath,amssymb}
\usepackage{newtxtext,newtxmath}  
\usepackage{microtype}

\usepackage{longtable,booktabs,array}
\usepackage{calc}
\usepackage{etoolbox}
\makeatletter
\patchcmd\longtable{\par}{\if@noskipsec\mbox{}\fi\par}{}{}
\makeatother
\usepackage{footnote}
\makesavenoteenv{longtable}

\usepackage{xcolor}
\usepackage{tikz}
\usepackage{pgfplots}
\pgfplotsset{compat=1.18}
\usepackage{graphicx}
\usepackage{float}

\usepackage{enumitem}

\usepackage[most]{tcolorbox}
\tcbset{
  resultbox/.style={colback=gray!5, colframe=gray!50!black, fonttitle=\bfseries,
    boxrule=0.5pt, arc=2pt, left=6pt, right=6pt, top=4pt, bottom=4pt},
  principlebox/.style={colback=blue!3, colframe=blue!40!black, fonttitle=\bfseries,
    boxrule=0.5pt, arc=2pt, left=6pt, right=6pt, top=4pt, bottom=4pt},
}

\usepackage[hyphens]{url}
\usepackage{hyperref}
\hypersetup{
  colorlinks=true,
  linkcolor=blue!70!black,
  citecolor=green!50!black,
  urlcolor=blue!70!black,
  pdftitle={The Kitchen Loop: User-Spec-Driven Development for a Self-Evolving Codebase},
  pdfauthor={Yannick Roy},
}
\usepackage{bookmark}

\newcommand{\roughly}{{\textraisebox{.17ex}{$\scriptstyle\sim$}}}
\makeatletter
\newcommand{\textraisebox}[2]{\leavevmode\raise#1\hbox{#2}}
\makeatother

\providecommand{\tightlist}{%
  \setlength{\itemsep}{0pt}\setlength{\parskip}{0pt}}

\title{The Kitchen Loop: User-Spec-Driven Development\\for a Self-Evolving Codebase}

\author{
  Yannick Roy\\
  \texttt{0xAgentKitchen@gmail.com}
}

\date{March 2026}

\begin{document}

\maketitle

\begin{abstract}
Code production is now a commodity; the bottleneck is \emph{knowing what to build} and \emph{proving it works}. We present the \textbf{Kitchen Loop},\footnote{Open-source: \url{https://github.com/0xagentkitchen/kitchenloop}} a framework for autonomous, self-evolving software built on a unified trust model: (1)~a \textbf{specification surface} enumerating what the product claims to support; (2)~\textbf{``As a User $\times$ 1000''}, where an LLM agent exercises that surface as a synthetic power user at ${\sim}1{,}000\times$ human cadence; (3)~\textbf{Unbeatable Tests}, ground-truth verification the code author cannot fake; and (4)~\textbf{Drift Control}, continuous quality measurement with automated pause gates. We validate across two production systems over 285+ iterations, producing 1,094+ merged pull requests with zero regressions detected by the regression oracle (methodology in Section~\ref{sec:methodology-what-zero-regressions-means}). We observe emergent properties at scale: multi-iteration self-correction chains, autonomous infrastructure healing, and monotonically improving quality gates. The primitives are not new; our contribution is their composition into a production-tested system with the operational discipline that makes long-running autonomous evolution safe.
\end{abstract}

\vspace{1em}

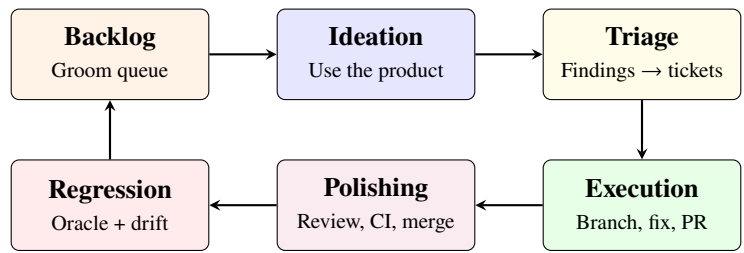
\begin{figure}[H]
\centering
\begin{tikzpicture}[
    phase/.style={draw, rounded corners=3pt, minimum width=2.6cm, minimum height=1.2cm, align=center, font=\small, fill=#1},
    arr/.style={->, thick, >=stealth}
]
\node[phase=orange!10] (backlog) at (0, 1.8) {\textbf{Backlog}\\{\scriptsize Groom queue}};
\node[phase=blue!10] (ideate) at (3.5, 1.8) {\textbf{Ideation}\\{\scriptsize Use the product}};
\node[phase=yellow!10] (triage) at (7, 1.8) {\textbf{Triage}\\{\scriptsize Findings $\rightarrow$ tickets}};
\node[phase=green!10] (execute) at (7, -0.2) {\textbf{Execution}\\{\scriptsize Branch, fix, PR}};
\node[phase=purple!8] (polish) at (3.5, -0.2) {\textbf{Polishing}\\{\scriptsize Review, CI, merge}};
\node[phase=red!8] (regress) at (0, -0.2) {\textbf{Regression}\\{\scriptsize Oracle + drift}};

\draw[arr] (backlog) -- (ideate);
\draw[arr] (ideate) -- (triage);
\draw[arr] (triage) -- (execute);
\draw[arr] (execute) -- (polish);
\draw[arr] (polish) -- (regress);
\draw[arr] (regress) -- (backlog);
\end{tikzpicture}
\caption{The Kitchen Loop: a six-phase autonomous improvement cycle.}
\label{fig:six-phase-overview}
\end{figure}

\newpage
\section{Core Contributions}\label{sec:core-contributions}

This paper makes four claims:

\begin{enumerate}
\def\labelenumi{\arabic{enumi}.}
\item
  \textbf{``As a User'' vs.~task-completion.} Agentic systems should not just close tickets --- they should systematically exercise a product's specification surface the way a user would. The Kitchen Loop is a production-tested framework for this, grounded in synthetic user journeys rather than isolated issues.
\item
  \textbf{The unified trust model.} Autonomous evolution requires a unified trust model. Three components interlock to make it safe: a \emph{specification surface} (what the product claims to do), \emph{unbeatable tests} (ground-truth verification the author can't fake), and a regression oracle with \emph{drift control} and automatic pause gates.
\item
  \textbf{Unbeatable tests.} Correctness requires adversarial, multi-model review. Implementer-written tests are necessary but insufficient --- in our deployments, 38 passing unit tests coexisted with complete feature failure. The Kitchen Loop enforces\\
  correctness through adversarial UAT gates (sealed test cards, fresh evaluator, zero context) and mandatory cross-model review: every PR is challenged by independent agents (Codex, Gemini, CodeRabbit) before merge. No output is accepted as-is from the model that wrote it.
\item
  \textbf{Bounded production evidence.} Across two production systems and 285+ iterations, the Kitchen Loop produced 1,094+ merged pull requests with zero regressions detected by the regression oracle (methodology in Section 4.1), monotonically improving quality gates (76--91\% → 100\%), and a cost of \roughly\$0.38 per merged PR.
\end{enumerate}

\vspace{0.8em}
\textbf{Key definitions used throughout:}

{\def\LTcaptype{none} 
\begin{longtable}[]{@{}
  >{\raggedright\arraybackslash}p{(\linewidth - 2\tabcolsep) * \real{0.3529}}
  >{\raggedright\arraybackslash}p{(\linewidth - 2\tabcolsep) * \real{0.6471}}@{}}
\toprule\noalign{}
\begin{minipage}[b]{\linewidth}\raggedright
Term
\end{minipage} & \begin{minipage}[b]{\linewidth}\raggedright
Definition
\end{minipage} \\
\midrule\noalign{}
\endhead
\bottomrule\noalign{}
\endlastfoot
\textbf{Specification surface} & The enumerable set of capabilities a product claims to support --- the input to coverage-exhaustion \\
\textbf{Unbeatable test} & A test that verifies outcomes against ground truth that the code author cannot fake \\
\textbf{Regression oracle} & A repeatable, bounded test that answers ``is the system at least as good as before?'' \\
\textbf{Coverage-exhaustion mode} & An operating regime where the agent systematically exercises every combination in the specification surface until coverage gaps approach zero \\
\textbf{UAT gate} & Adversarial user-acceptance testing by a fresh evaluator with zero implementation context \\
\textbf{Drift control} & Continuous measurement of quality trends with automated pause gates that halt the loop when metrics degrade \\
\end{longtable}
}

\newpage
\section{Executive Summary (TL;DR)}\label{sec:executive-summary-tldr}

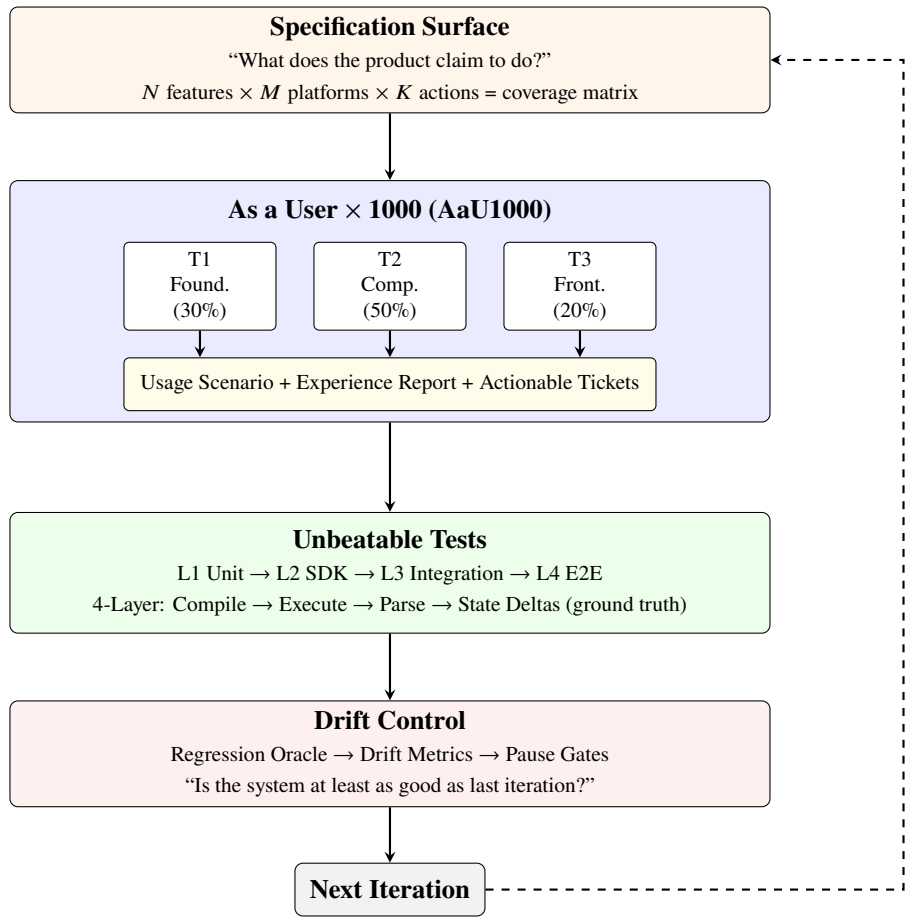
\begin{figure}[H]
\centering
\begin{tikzpicture}[
    block/.style={draw, rounded corners=3pt, minimum width=10cm, align=center, font=\small, fill=#1},
    arr/.style={->, thick, >=stealth},
    tier/.style={draw, rounded corners=2pt, minimum width=2cm, minimum height=0.8cm, align=center, font=\scriptsize, fill=white}
]
\node[block=orange!8, minimum height=1.2cm] (spec) at (0, 11) {\textbf{Specification Surface}\\{\scriptsize ``What does the product claim to do?''}\\{\scriptsize $N$ features $\times$ $M$ platforms $\times$ $K$ actions = coverage matrix}};

\node[block=blue!8, minimum height=3.2cm] (aau) at (0, 7.8) {};
\node[font=\small\bfseries] at (0, 9) {As a User $\times$ 1000 (AaU1000)};
\node[tier] (t1) at (-2.5, 8) {T1\\Found.\\(30\%)};
\node[tier] (t2) at (0, 8) {T2\\Comp.\\(50\%)};
\node[tier] (t3) at (2.5, 8) {T3\\Front.\\(20\%)};
\node[draw, rounded corners=2pt, minimum width=7cm, minimum height=0.7cm, fill=yellow!10, font=\scriptsize] (out) at (0, 6.7) {Usage Scenario + Experience Report + Actionable Tickets};

\draw[arr] (spec.south) -- ++(0, -0.4) -- (aau.north);

\draw[arr] (t1.south) -- (out.north -| t1.south);
\draw[arr] (t2.south) -- (out.north);
\draw[arr] (t3.south) -- (out.north -| t3.south);

\node[block=green!8, minimum height=1.6cm] (tests) at (0, 4.2) {\textbf{Unbeatable Tests}\\{\scriptsize L1 Unit $\rightarrow$ L2 SDK $\rightarrow$ L3 Integration $\rightarrow$ L4 E2E}\\{\scriptsize 4-Layer: Compile $\rightarrow$ Execute $\rightarrow$ Parse $\rightarrow$ State Deltas (ground truth)}};

\draw[arr] (aau.south) -- ++(0, -0.4) -- (tests.north);

\node[block=red!6, minimum height=1.2cm] (drift) at (0, 1.8) {\textbf{Drift Control}\\{\scriptsize Regression Oracle $\rightarrow$ Drift Metrics $\rightarrow$ Pause Gates}\\{\scriptsize ``Is the system at least as good as last iteration?''}};

\draw[arr] (tests.south) -- ++(0, -0.4) -- (drift.north);

\node[draw, rounded corners=3pt, minimum width=2.5cm, minimum height=0.7cm, fill=gray!10, font=\small\bfseries] (next) at (0, 0) {Next Iteration};

\draw[arr] (drift.south) -- ++(0, -0.3) -- (next.north);

\draw[arr, dashed] (next.east) -- ++(5.5,0) |- (spec.east);
\end{tikzpicture}
\caption{The unified trust model: each iteration flows through the full verification stack before proceeding.}
\label{fig:trust-model-overview}
\end{figure}

\begin{description}[style=nextline, leftmargin=1.5em, font=\bfseries]
\item[The method] An LLM agent uses the product as a synthetic power user at ${\sim}1{,}000\times$ human cadence against its specification surface and beyond, validates through unbeatable tests, and controls drift and regression before accepting the work to ensure the product self-evolves in the right direction (Figure~\ref{fig:trust-model-overview}).
\end{description}

\begin{description}[style=nextline, leftmargin=1.5em, font=\bfseries]
\item[The architecture] A six-phase loop (Figure~\ref{fig:six-phase-overview}) with automated drift control and pause gates.

\item[The results] Seven weeks, two production systems:
\end{description}

\begin{tcolorbox}[resultbox, title=Kitchen Loop Results]
\centering
\renewcommand{\arraystretch}{1.3}
\begin{tabular}{@{}r@{\;=\;}l@{\qquad}r@{\;=\;}l@{\qquad}r@{\;=\;}l@{}}
Iterations & 285+ & PRs merged & 1{,}094+ & Tickets & 700+ \\
Tests & 13{,}000+ & Regressions & \textbf{0} & Cost/PR & ${\sim}$\$0.38 \\
\end{tabular}

\vspace{6pt}
\begin{tabular}{@{}r@{\quad}l@{}}
\toprule
Quality gates & L1 100\%, L2 100\%, L3 100\% (from 76--91\%) \\
Canary escapes (Tier 1) & 0 across all 163 signal iterations \\
Monthly cost & ${\sim}$\$350 (flat-rate subscriptions, both systems) \\
Iteration speed & 5 min (signals) to 80--230 min (execution) \\
Production incidents & \textbf{0} \\
\bottomrule
\end{tabular}
\end{tcolorbox}

\section{The Post-Commodity Code Thesis}\label{sec:post-commodity-code-thesis}

\subsection{Code Is No Longer the Hard Part}\label{sec:code-is-no-longer-the-hard-part}

LLM-based coding agents can produce functional code at a rate that renders the writing step non-limiting. Robbes et al.~(2026) find that 15--23\% of mature open-source projects adopted coding agents within just nine months of tool availability \cite{robbes2026}. A senior engineer's value is shifting from writing code to knowing \emph{which} code to write, \emph{why}, and \emph{how to prove it's correct}.

Yet the productivity story is more nuanced than adoption rates suggest. Becker et al.~(2025) conducted the most rigorous randomized controlled trial to date: experienced open-source developers using AI copilots were 19\% \emph{slower} than those working without AI, contradicting both developer self-estimates (+20\% speedup) and expert forecasts (+38\% speedup) \cite{becker2025}. He et al.~(2025) provide causal evidence that Cursor adoption produces a large but transient velocity boost (+281\% lines added in month one, dissipating by month three) accompanied by persistent quality degradation: +30\% static analysis warnings, +42\% cognitive complexity, and +7\% code duplication \cite{he2025}. The accumulated complexity feeds back into velocity --- every doubling of complexity reduces future velocity by \roughly64.5\% \cite{he2025}. The Kitchen Loop can be read as a response to both findings: by shifting emphasis from code generation to specification and unbeatable verification, it addresses the quality degradation that erodes velocity gains and the lack of structural verification that leaves AI-assisted development slower than expected.

The same shift is happening to code review. Automated review tools can identify bugs, style violations, and security issues at a pace no human reviewer can match. The human reviewer's role shifts from ``find the bug'' to ``judge whether this code serves the product's intent.''

This creates a new bottleneck: \textbf{specification and verification}. The hard problems are now:
\begin{itemize}[nosep]
\item What should the product do? (Specification)
\item Does the product actually do it? (Verification)
\item Is the product getting better or worse over time? (Drift control)
\end{itemize}

\subsection{The Turing Test for Code}\label{sec:turing-test-for-code}

Alan Turing's insight was that you don't need to understand how a machine thinks --- you need a test rigorous enough that passing it constitutes sufficient evidence of capability. The machine can remain a black box. The test is the arbiter.

We apply the same principle to AI-generated code. When an LLM agent writes a function, a connector, or a full feature, you face a choice:

\begin{itemize}
\tightlist
\item
  \textbf{Option A}: Read every line, understand every branch, verify the logic manually. This scales poorly when the agent produces thousands of lines per day.
\item
  \textbf{Option B}: Build tests so rigorous that if the code passes them, you can \emph{trust} it works --- without needing to audit every line. The code becomes a black box. The tests are the Turing Test.
\end{itemize}

Empirical evidence supports both options coexisting in practice. Huang et al.~(2025) find that 69\% of professional developers carefully review every agentic change and 75\% read every line of AI-generated code --- but crucially, developers working on \emph{unfamiliar} tasks let agents drive implementation while monitoring program \emph{outputs} rather than reviewing code line-by-line \cite{huang2025}. This validates Option B for the regime where the Kitchen Loop operates: autonomous agents on structured tasks with rigorous output verification.

The urgency for Option B is underscored by a growing QA crisis in AI-assisted development. Fawzy et al.~(2025) find that 36\% of practitioners using AI code generation skip quality assurance entirely, 18\% place uncritical trust in AI output, and 10\% delegate QA back to the same AI that wrote the code \cite{fawzy2025}. The result is predictable: 68\% of practitioners characterize the output as ``fast but flawed'' \cite{fawzy2025}. Voluntary QA discipline is demonstrably insufficient; structural enforcement --- tests that \emph{cannot} be skipped --- is the only reliable solution.

Option B requires a specific kind of test: not unit tests written by the same agent that wrote the code (the agent could pass its own tests trivially), but \emph{end-to-end verification against real-world state}. For a DeFi SDK, this means executing transactions on chain forks and verifying balance changes. For a signal platform, this means cross-referencing claims against live data APIs. For a web application, this means browser automation against a real rendering engine. For any product, it means testing at the boundary where the software meets reality.

We call these \textbf{unbeatable tests} --- tests that verify outcomes against ground truth that the code author cannot fake. The test doesn't care \emph{how} the code achieves the result --- only that it does.

It is of utmost importance to be aware of the weaknesses of LLMs when it comes to tests. There is a fallacy to believe that LLMs are so good at writing code, that they should be equally good at writing tests. They aren't. They often obsess over the test itself forgetting that passing the test is not the goal in itself. That obsession to pass the test often leads to cheating, modifying the test or even writing side scripts nailing the test, leaving the actual codebase untested.

\section{Related Work}\label{sec:related-work}

The Kitchen Loop exists within a rapidly evolving landscape of agentic software engineering. We position our contributions against four categories of related work.

\subsection{Autonomous Coding Agents}\label{sec:autonomous-coding-agents}

\textbf{SWE-agent} (Yang et al., 2024) and \textbf{AutoCodeRover} (Zhang et al., 2024) demonstrated that LLM agents can resolve GitHub issues autonomously by navigating codebases, editing files, and running tests. \textbf{OpenHands} (Wang et al., 2024) and \textbf{Devin} (Cognition, 2024) extended this to multi-step workflows including environment setup, debugging, and deployment. \textbf{RepairAgent} (Bouzenia et al., 2024) focused specifically on automated program repair with LLM-driven fault localization.

These systems operate in \textbf{task-completion mode}: given an issue, produce a patch. The Kitchen Loop operates in \textbf{coverage-exhaustion mode}: given a specification surface, systematically exercise every combination until the gap between spec and reality approaches zero. The unit of work is not ``resolve this issue'' but ``attempt this user scenario end-to-end and document everything that breaks.'' Task completion is a component of the loop (the Execute phase), not the loop itself.

\subsection{Self-Improving and Self-Evolving Agents}\label{sec:self-improving-and-self-evolving-agents}

\textbf{SICA} (``A Self-Improving Coding Agent'', arXiv 2504.15228, 2025) introduced agents that edit their own prompts and tools based on execution feedback, achieving self-improvement without human intervention. \textbf{AlphaEvolve} (DeepMind, 2025) demonstrated repository-scale code evolution using LLMs guided by automated evaluators. \textbf{Confucius SDK} proposed a meta-agent build-test-improve loop for SDK development.

The Kitchen Loop shares the self-improvement property (Section 13) --- it discovers its own infrastructure bugs and updates its own skill files. The key difference is that self-improvement in the Kitchen Loop is \emph{anchored to a specification surface and regression oracle}, not to an optimization objective. The loop doesn't optimize for a metric; it converges toward a specification. This prevents the ``goodharting'' failure mode where an agent optimizes a proxy metric while the product degrades in unmeasured dimensions.

\subsection{Long-Running and Looping Patterns}\label{sec:long-running-and-looping-patterns}

\textbf{Ralph Wiggum loops} (Huntley, 2025; snarktank/ralph) demonstrated that autonomous agents can produce 1,000+ commits overnight in a continuous loop. Multi-agent SE frameworks like \textbf{MetaGPT} (Hong et al., 2024), \textbf{ChatDev} (Qian et al., 2024), and \textbf{ALMAS} assigned different agent roles (PM, architect, coder, tester) to simulate a software team.

The Kitchen Loop draws from this lineage but adds three structural elements that Ralph-style loops lack: (1) a \textbf{production-parity ticket management layer} (Linear) where human and AI tickets are treated identically, preventing the loop from diverging from human priorities; (2) \textbf{automated drift control} with pause gates that halt the loop when quality degrades; and (3) the \textbf{three-tier strategy model} (Foundation/Composition/Frontier) that ensures balanced coverage rather than random or greedy scenario selection.

\subsection{Spec-Driven and Verification-Focused Approaches}\label{sec:spec-driven-and-verification-focused-approaches}

\textbf{SpecRover} (Ruan et al., 2024) used natural-language specifications to guide automated debugging. \textbf{AgileGen} and \textbf{Augmented Agile} explored LLM-assisted requirements engineering and test generation from user stories. \textbf{USEagent} proposed user-story-driven end-to-end testing.

The Kitchen Loop's AaU1000 method is closest to USEagent's vision but differs in scope and trust model. USEagent generates tests from user stories; AaU1000 generates \emph{usage scenarios} from specification surfaces and validates them through \emph{unbeatable tests} (4-layer verification against ground truth). The distinction matters: a test generated from a user story proves the story is implemented; a usage scenario validated through state-delta verification proves the product \emph{actually works} in the way a real user would experience it.

\subsection{Vibe Coding and the QA Crisis}\label{sec:vibe-coding-and-the-qa-crisis}

Fawzy et al.~(2025) find a ``speed-quality paradox'' across 101 practitioner sources: 62\% of vibe coders cite speed as their primary motivation, yet 68\% characterize output as ``fast but flawed'' \cite{fawzy2025}. Huang et al.~(2025) confirm that professional developers reject the vibe coding paradigm, carefully controlling agents through strategic plans --- averaging only 2.1 execution steps per prompt despite plans spanning 70+ steps \cite{huang2025}. Their task suitability taxonomy identifies agents as unsuitable for business logic, domain knowledge, complex reasoning, and security-critical code --- precisely the categories where the Kitchen Loop's spec surface and UAT gate provide structural guardrails.

\subsection{Benchmark Contamination, Adoption, and Quality Evidence}\label{sec:benchmark-contamination-adoption-quality}

The dominant evaluation benchmark, SWE-Bench, suffers from data contamination: Liang et al.~(2025) show a 23-percentage-point accuracy gap between memorized and novel repositories \cite{liang2025}, while Thai et al.~(2025) find GPT-5 scores 65\% on single-issue tasks but only 21\% on multi-file evolution tasks \cite{thai2025}. Meanwhile, adoption is accelerating --- Robbes et al.~(2026) find 15--23\% of mature open-source projects adopted coding agents within nine months \cite{robbes2026} --- but quality lags: 90.6\% of agent-authored PRs receive zero human review \cite{rahman2026}, and \roughly40\% of Copilot-generated code contains security vulnerabilities \cite{gao2025}. The Kitchen Loop addresses both problems: it rejects benchmark-style evaluation in favor of verification against live product state, and interposes adversarial verification between generation and merge.

\subsection{Multi-Agent Design Patterns}\label{sec:multi-agent-design-patterns}

Cai et al.~(2025) systematically review 94 LLM-based multi-agent systems for software engineering, identifying 16 design patterns ranked by adoption frequency \cite{cai2025}. The Kitchen Loop composes seven of these patterns into a single integrated loop: Role-Based Cooperation (six-phase loop with distinct roles per phase), Self-Reflection (Polish and Regress phases), Cross-Reflection (UAT Gate), Debate-Based Cooperation (Discussion Manager), Voting-Based Cooperation (multi-model tribunal), Tool-Agent Registry (skills directory), and Agent Evaluator (regression oracle). Notably, three of these --- Debate-Based Cooperation (4.3\% adoption), Voting-Based Cooperation (3.2\%), and Agent Evaluator (3.2\%) --- are among the least adopted patterns in the literature \cite{cai2025}, suggesting the Kitchen Loop operationalizes architectural ideas the field has not yet widely implemented.

\subsection{Our Differentiation}\label{sec:our-differentiation}

\begin{table}[ht]
\centering
\caption{Comparison of agentic software engineering approaches.}
\label{tab:differentiation}
\footnotesize
\renewcommand{\arraystretch}{1.25}
\begin{tabular}{@{}l l l l l@{}}
\toprule
& \textbf{Task-Completion} & \textbf{Ralph Loops} & \textbf{Self-Improving} & \textbf{Kitchen Loop} \\
\midrule
Unit of work       & Issue $\to$ Patch       & Commit          & Optimization step   & Scenario $\to$ experience report \\
Stopping cond.     & Issue resolved           & Time/count      & Metric plateau      & Specification exhausted \\
Coverage           & Reactive (given issues)  & Random/greedy   & Objective-guided    & Three-tier (F/C/F) \\
Quality gate       & CI passes                & CI + typecheck  & Evaluator function  & 4-layer + multi-model tribunal \\
Drift control      & None                     & Basic (progress log) & Implicit (metrics)  & Explicit (oracle + pause gates) \\
Human role         & PR review                & Manual oversight & None               & Shared backlog, ticket parity \\
Self-improvement   & None                     & Manual (AGENTS.md) & Core feature      & Anchored to spec + oracle \\
\bottomrule
\end{tabular}
\end{table}

From our analysis, three distinct operating regimes for agentic software engineering emerge:

\begin{table}[ht]
\centering
\caption{Three operating regimes for agentic software engineering.}
\label{tab:regimes}
\small
\renewcommand{\arraystretch}{1.25}
\begin{tabular*}{\textwidth}{@{\extracolsep{\fill}}l l l l l@{}}
\toprule
\textbf{Regime} & \textbf{Unit of Work} & \textbf{Stopping Cond.} & \textbf{Verification Target} & \textbf{Failure Mode} \\
\midrule
Task-completion      & Issue          & Issue resolved     & Patch correctness     & Local fix / global mismatch \\
Metric-optimization  & Objective step & Metric plateau     & Proxy metric          & Goodharting \\
Coverage-exhaustion  & User scenario  & Surface exhaustion & User-visible behavior & Drift / incomplete surface \\
\bottomrule
\end{tabular*}
\end{table}

The Kitchen Loop operates in the coverage-exhaustion regime. The regime determines the system's failure mode: task-completion risks local fixes that break global behavior; metric-optimization risks Goodharting on proxy measures; coverage-exhaustion risks drift or incomplete specification surfaces, which the drift control mechanisms in Section~\ref{sec:controlling-regression-and-drift} are designed to address.

\section{The ``As a User x 1000'' Method (AaU1000)}\label{sec:as-a-user-x-1000-method-aau1000}

\subsection{The Core Insight}\label{sec:core-insight}

The most reliable signal about whether a feature works is a real usage attempt --- not a unit test, not a code review, but someone exercising the product end-to-end. The problem: human usage attempts are slow and expensive. The AaU1000 method replaces the ``someone'' with an LLM agent that:

\begin{enumerate}
\def\labelenumi{\arabic{enumi}.}
\tightlist
\item
  \textbf{Selects a realistic usage scenario} derived from the product's specification surface
\item
  \textbf{Attempts it as a real user would} --- writing code, running it, observing what breaks
\item
  \textbf{Documents failures as actionable tickets} --- not vague observations but precise bug reports with reproduction steps, root cause hypotheses, and file-level pointers
\item
  \textbf{Fixes those failures immediately} --- implementing the fix, writing tests, and shipping a PR
\item
  \textbf{Watches for regression and drift} --- verifying nothing else broke and the codebase is not getting worse
\end{enumerate}

And then does it again. At whatever cadence the infrastructure allows.

\subsection{Quantifying the 1000x}\label{sec:quantifying-the-1000x}

The ``1000x'' is an order-of-magnitude claim, not a precise multiplier. We ground it empirically:

\textbf{Single-thread velocity}: A senior engineer realistically ships \roughly15-25 merged PRs per month (bug fixes, features, tests). Each Kitchen Loop instance runs single-threaded per product. The strategy framework produced 728+ merged PRs in \roughly5 weeks (\roughly145/week); the signal platform produced 366 in 17 days (\roughly150/week). Per-system, this is a \textbf{\roughly24-48x single-thread throughput increase} depending on domain iteration speed (the signal platform iterates \roughly25x faster than the SDK). Combined across both systems running concurrently, total output was 1,094+ merged PRs.

\textbf{Scenario throughput}: The DeFi strategy framework's loop completes one full usage scenario (ideate $\rightarrow$ implement $\rightarrow$ test $\rightarrow$ fix $\rightarrow$ regress) in 80-230 minutes. A human attempting the same scenario takes 1-3 days. This is a \textbf{7-25x per-scenario speedup}.

\textbf{Parallelization potential}: The loop currently runs single-threaded per product. Running N loops on N products (as demonstrated with SDK + Edge concurrently) scales linearly. With parallel execution infrastructure (dynamic port allocation, containerized test environments), a single product could run N loops exploring different specification regions simultaneously. The 1000x represents the \emph{achievable ceiling} with moderate parallelization --- not the current single-thread reality.

\subsection{Spec-Driven, Not Spec-Speculative}\label{sec:spec-driven-not-spec-speculative}

The method is scoped to \textbf{what exists and is expected to work today} --- not speculative feature engineering, benchmarking, or user research.

\subsection{The ``Obviously Missing'' Signal}\label{sec:obviously-missing-signal}

The most important class of findings is the \textbf{``obviously missing'' signal}: features any competent user would expect, but that don't work. These are objectively verifiable failures --- a function returning \texttt{AttributeError}, a missing configuration entry, a gas cap blocking all transactions on a chain. The loop discovers verifiable failures, not speculative issues.

\subsection{The Three-Tier Strategy Model}\label{sec:three-tier-strategy-model}

The Kitchen Loop formalizes a \textbf{three-tier scenario generation model} that ensures balanced coverage across the product's maturity spectrum:

\begin{figure}[h]
\centering
\begin{tikzpicture}
\def\rowa{0}
\def\rowb{-1.4}
\def\rowc{-2.8}

\fill[green!35, rounded corners=2pt] (0,\rowa+0.4) rectangle (3.0,\rowa-0.4);
\fill[blue!25, rounded corners=2pt]  (0,\rowb+0.4) rectangle (5.0,\rowb-0.4);
\fill[orange!30, rounded corners=2pt](0,\rowc+0.4) rectangle (2.0,\rowc-0.4);

\node[anchor=east, font=\bfseries\large] at (-0.3, \rowa) {T1};
\node[anchor=east, font=\bfseries\large] at (-0.3, \rowb) {T2};
\node[anchor=east, font=\bfseries\large] at (-0.3, \rowc) {T3};

\node[anchor=west, font=\bfseries] at (0.2, \rowa) {30\%};
\node[anchor=west, font=\bfseries] at (0.2, \rowb) {50\%};
\node[anchor=west, font=\bfseries] at (0.2, \rowc) {20\%};

\node[anchor=west, font=\small] at (5.4, \rowa) {\textbf{Foundation} --- ``Does the basic stuff work perfectly?''};
\node[anchor=west, font=\small] at (5.4, \rowb) {\textbf{Composition} --- ``What breaks when we combine things?''};
\node[anchor=west, font=\small] at (5.4, \rowc) {\textbf{Frontier} --- ``What's missing for the next generation?''};
\end{tikzpicture}
\end{figure}

\textbf{Tier 1 --- Foundation (30\%)} exercises a single feature on a single platform or configuration. One integration, one action, happy path. These scenarios should be trivially achievable by a new user in a few minutes. If anything goes wrong, it is a critical regression. Foundation iterations maintain the baseline: the easy stuff must always be bulletproof.

\textbf{Tier 2 --- Composition (50\%)} combines two or more features in creative ways. Multi-service flows, indicator-driven behavior, multi-step workflows, cross-platform configuration stress tests. The goal is to find bugs at the \emph{seams} between components that pass individually but fail in combination. This is where the majority of novel bug discovery happens, because the combinatorial space of feature pairs and triples is far larger than the space of individual features.

\textbf{Tier 3 --- Frontier (20\%)} deliberately reaches beyond the product's current capabilities. The deliverable shifts from ``working scenario'' to ``gap analysis'': what would you need to build this? What's missing? What would it unlock? The experience report emphasizes specific missing features, ordered by implementation effort and user value.

\subsection{The Self-Expanding Property}\label{sec:self-expanding-property}

The three tiers form a \textbf{self-reinforcing growth cycle}. This is the ``secret sauce'' that transforms a random agent into a strategic engineer:

\begin{tcolorbox}[resultbox]
\ttfamily\small
\begin{tabular*}{\linewidth}{@{\quad} l @{\quad} l @{\extracolsep{\fill}} r @{}}
\multicolumn{3}{@{}l}{Iteration N: \quad Features = \{A, B\}} \\[3pt]
T1 (Foundation):  & test A, test B                                    & (2 scenarios) \\
T2 (Composition): & test A+B, test B+A                                & (2 scenarios) \\
T3 (Frontier):    & ``I need C to do A+C'' $\rightarrow$ gap analysis & (1 gap report) \\[6pt]
\multicolumn{3}{@{}l}{--- C gets built ---} \\[6pt]
\multicolumn{3}{@{}l}{Iteration N+k: \quad Features = \{A, B, C\}} \\[3pt]
T1 (Foundation):  & test A, test B, test C          & (3 scenarios) \\
T2 (Composition): & test A+B, A+C, B+C, A+B+C      & (4+ scenarios) \\
T3 (Frontier):    & ``Now I want D to do A+C+D''    & (1 gap report) \\
\end{tabular*}
\end{tcolorbox}

Foundation grows linearly with each new feature. Composition grows \textbf{superlinearly} --- each new feature can be combined with every existing feature and pair. Frontier pushes the boundary outward. The loop evolves the product by systematically testing what exists and identifying the next most valuable capability.

\begin{tcolorbox}[principlebox, title=Design Principles]
\begin{description}[leftmargin=2.5em, style=nextline]
\item[P1. Ground-Truth Verification] Trust in AI-generated code should be proportional to how ground-truth-verifiable the test outcomes are, not to test coverage percentage.
\item[P2. Weakest-Evaluator] A user test is only trustworthy if a minimally capable evaluator can execute it. (\S4.8)
\item[P3. Spec-Anchored Improvement] Self-improving agents should optimize toward specification satisfaction, not proxy metrics. (\S2.2, \S3.5)
\item[P4. Drift-Before-Failure] Autonomous loops need continuous trend monitoring, not only binary quality gates. (\S5.5)
\end{description}
\end{tcolorbox}

\section{Unbeatable Tests: The Multi-Tier QA Framework}\label{sec:unbeatable-tests-the-multi-tier-qa-framework}

\subsection{Methodology: What ``Zero Regressions'' Means}\label{sec:methodology-what-zero-regressions-means}

Throughout this paper, we report ``zero regressions introduced by loop-merged code.'' This claim requires precise definition to be falsifiable:

\begin{itemize}
\tightlist
\item
  \textbf{Detection method}: Every iteration ends with an automated regression oracle run (Section 5.2). For the DeFi strategy framework, this means executing demo strategies on Anvil chain forks and verifying 4-layer state deltas. For the signal platform, this means running the full quality-gate pipeline (structural, factual, temporal, cognitive) plus anti-signal canaries.
\item
  \textbf{Measurement window}: Continuous --- every iteration, not sampled. The oracle runs after every merge, not on a periodic schedule.
\item
  \textbf{Scope}: The claim covers regressions \emph{detected by the oracle} in code \emph{merged through the loop's own process}. It does not claim the codebase is bug-free, nor that latent regressions undetectable by the oracle do not exist. The oracle's coverage is bounded by its test suite (10,913 unit tests, 62 demo strategies, 77 signal verifiers).
\item
  \textbf{Exclusions}: Environmental failures (API timeouts, chain fork instability) are distinguished from code regressions by correlation with recent merges. A failure that reproduces on the pre-merge commit is environmental; a failure introduced by a specific PR is a regression.
\end{itemize}

This definition makes the claim auditable: any third party with access to the oracle suite and git history can reproduce the measurement.

\subsection{Why Tests Are the New Competitive Advantage}\label{sec:why-tests-are-the-new-competitive-advantage}

Gao et al.~(2025) find functional bugs in 78\% of studies on AI-generated code, with \roughly40\% of Copilot code containing security vulnerabilities \cite{gao2025}. Dominant benchmarks (HumanEval, MBPP) rely on Pass@k metrics that overlook semantic correctness \cite{gao2025}, and Liang et al.~(2025) show SWE-Bench performance is inflated by data contamination (76\% accuracy on memorized paths vs.~53\% on novel repositories) \cite{liang2025}. Unbeatable tests sidestep both failure modes: they verify against live product state that changes with every deployment, making both contamination and pass-rate gaming irrelevant.

This requires tests at multiple tiers, each catching a different class of defect that lower tiers miss.

\subsection{The 4-Level Testing Pyramid}\label{sec:4-level-testing-pyramid}

\begin{figure}[h]
\centering
\begin{tikzpicture}[
    level/.style={draw, rounded corners=2pt, minimum height=1.1cm, text width=#1, align=center, font=\small},
    label/.style={font=\small\itshape, anchor=west}
]
\node[level=5.5cm, fill=blue!15] (l4) at (0, 3.3) {\textbf{Level 4: E2E Scenario Tests}\\{\scriptsize Scenario $\rightarrow$ Actions $\rightarrow$ Execution $\rightarrow$ State}};
\node[level=7.5cm, fill=blue!10] (l3) at (0, 2.0) {\textbf{Level 3: Integration Tests}\\{\scriptsize Compile, execute, verify against ground truth}};
\node[level=9.5cm, fill=gray!10] (l2) at (0, 0.7) {\textbf{Level 2: API/Adapter Tests}\\{\scriptsize Individual method tests with real dependencies}};
\node[level=11.5cm, fill=gray!5] (l1) at (0, -0.6) {\textbf{Level 1: Unit Tests}\\{\scriptsize Isolated, mocked, fast --- pure function correctness}};
\node[label] at (3.3, 3.3) {Full user journey};
\node[label] at (4.3, 2.0) {Real execution};
\node[label] at (5.3, 0.7) {API contracts};
\node[label] at (6.3, -0.6) {Logic validation};
\end{tikzpicture}
\end{figure}

\begin{table}[ht]
\centering
\small
\renewcommand{\arraystretch}{1.25}
\begin{tabular*}{\textwidth}{@{\extracolsep{\fill}} l l l l l @{}}
\toprule
\textbf{Level} & \textbf{What It Tests} & \textbf{Speed} & \textbf{Trust Level} & \textbf{Written By} \\
\midrule
\textbf{L1} & Isolated logic, calculations  & Fast (ms)  & Low --- proves logic, not integration & Agent or human \\
\textbf{L2} & API methods, adapters         & Medium (s) & Medium --- proves API contracts      & Agent or human \\
\textbf{L3} & Full execution pipeline       & Slow (min) & High --- proves real-world behavior  & Loop + agent \\
\textbf{L4} & Complete user journeys        & Slowest    & Highest --- proves product works     & Loop (AaU1000) \\
\bottomrule
\end{tabular*}
\end{table}

\textbf{The critical insight}: L1 and L2 are \emph{necessary but not sufficient}. A function can pass all its unit tests and still fail in production because the unit tests don't exercise the real execution environment. \textbf{L3 and L4 are the unbeatable tests} --- they verify against ground truth (real state, real APIs, real execution) that the code author cannot fake.

\subsection{The 4-Layer Verification Pattern}\label{sec:4-layer-verification-pattern}

Every L3 integration test implements \textbf{four verification layers}, each catching a different class of defect. We illustrate with the DeFi strategy framework (Case Study A, Section~\ref{sec:case-study-a-defi-strategy-framework}); other domains substitute their own verification layers (e.g., browser automation for web apps, API contract testing for backends --- see Section~\ref{sec:generalization-making-the-kitchen-loop-portable} for adaptation examples).

\begin{table}[ht]
\centering
\small
\renewcommand{\arraystretch}{1.25}
\begin{tabular*}{\textwidth}{@{\extracolsep{\fill}} c l l l @{}}
\toprule
\textbf{Layer} & \textbf{Name} & \textbf{What It Catches} & \textbf{How} \\
\midrule
\textbf{1} & Compilation      & Wrong params, missing config, type errors      & Build/compile the action, assert success \\
\textbf{2} & Execution        & Runtime failures, permission errors, timeouts  & Execute against real env, assert success \\
\textbf{3} & Output Parsing   & Wrong decoding, missing fields, malformed data & Parse output, assert expected data extracted \\
\textbf{4} & State Deltas     & Silent failures, partial exec, wrong outcomes  & Measure state before/after, assert exact deltas \\
\bottomrule
\end{tabular*}
\end{table}

A test that only compiles is \textbf{incomplete}. A test that compiles and executes but doesn't check state deltas is \textbf{dangerously incomplete} --- it could silently succeed while doing the wrong thing. Layer 4 is what makes the test unbeatable: it verifies the \emph{outcome}, not just the \emph{execution}.

For failure-mode tests: 3 layers are required (compilation, execution, state deltas), and state conservation MUST be asserted (state before and after remain unchanged). The system must not silently lose assets when it fails.

\subsection{The Coverage Matrix: Exhaustive by Design}\label{sec:coverage-matrix-exhaustive-by-design}

The specification defines a coverage matrix: every combination of feature, platform, and action type that the product claims to support. The test suite's goal is to fill every cell in this matrix.

For a product with N features, M platforms, and K action types, the matrix has N x M x K cells. Each cell represents a claim: ``Feature X works on Platform Y for Action Z.'' Each empty cell is an untested claim --- a place where the product might silently fail.

The Kitchen Loop fills these cells systematically, prioritizing by risk:

\begin{enumerate}
\def\labelenumi{\arabic{enumi}.}
\tightlist
\item
  \textbf{P0}: Core features on primary platforms (table stakes)
\item
  \textbf{P1}: Core features on secondary platforms (breadth)
\item
  \textbf{P2}: Advanced features on primary platforms (depth)
\item
  \textbf{P3}: Edge cases and aggregators (completeness)
\end{enumerate}

No manual QA process can fill a matrix with hundreds or thousands of cells on a weekly cadence. The Kitchen Loop can, because it generates and runs tests at AI speed.

\subsection{Anti-Signal Canaries: Testing the Tests}\label{sec:anti-signal-canaries-testing-the-tests}

For systems where outputs are non-deterministic (signals, recommendations, generated content), the Kitchen Loop uses \textbf{anti-signal canaries}: intentionally crafted bad inputs injected alongside real ones to verify the quality gate catches what it should.

Four tiers of increasing deceptiveness:

\begin{table}[ht]
\centering
\footnotesize
\renewcommand{\arraystretch}{1.3}
\begin{tabular*}{\textwidth}{@{\extracolsep{\fill}} l l p{5.5cm} c l @{}}
\toprule
\textbf{Tier} & \textbf{Name} & \textbf{Description} & \textbf{Expected} & \textbf{Observed (iter 163)} \\
\midrule
\textbf{1} & Obviously Bad  & Glaring structural or factual errors any gate should catch                      & 100\%  & 100\% (0 escapes / 163 iters) \\
\textbf{2} & Shadow         & Factually true but stale, low-novelty, or below threshold                      & 50--80\%  & 100\% (from 33\% at iter 1, fixed iter 124) \\
\textbf{3} & Adversarial    & Real data with wrong conclusions --- fools deterministic and LLM checks         & 30--60\%  & 100\% (from 67\% at iter 1) \\
\textbf{4} & Mixed T/F      & Blend of valid and fabricated data in one signal --- partial-failure detection   & 20--50\%  & 100\% (added iter $\sim$100, fixed iter 124) \\
\bottomrule
\end{tabular*}
\end{table}

Additionally, \textbf{3 API degradation canaries} test resilience to partial external API failures (e.g., a data source returning errors or timeouts). These verify the quality gates degrade gracefully rather than producing false passes when dependencies fail.

The canary system provides a \textbf{known-bad baseline} that the regression phase can measure against. Tier 1 canary escapes are treated as a critical warning signal for operator review. The monotonic improvement across all tiers --- from partial catch rates in early iterations to 100\% across all 4 tiers by iteration 124 --- demonstrates that the loop's quality infrastructure improves alongside the product it protects.

A notable finding from the Edge deployment: Tier 2 catch rates were stuck at 33\% for 70+ iterations. Early attempts to improve them via the L4 LLM tribunal (GPT/Claude/Gemini judgment) produced non-deterministic results --- catching 0-2 canaries per iteration with no convergence. The durable fix came from encoding failure patterns as \emph{deterministic rules} (e.g., \texttt{STALE\_NARRATIVES}, \texttt{STALE\_CATALYSTS}), achieving 100\% catch rate that held for 40+ subsequent iterations. This validates Design Principle P1 (Ground-Truth Verification): for safety-critical gates, deterministic verification outperforms probabilistic LLM judgment.

\subsection{Multi-Model Review Tribunals}\label{sec:multi-model-review-tribunals}

For decisions that require judgment (architectural choices, ambiguous test results, code quality assessment), the Kitchen Loop uses \textbf{multi-model tribunals}: three independent AI reviewers evaluate the same artifact in parallel. Findings are synthesized with consensus classification:

\begin{itemize}
\tightlist
\item
  \textbf{Consensus} (all three agree): treated as a confirmed finding
\item
  \textbf{Majority} (two agree): treated as a likely finding, prioritized for action
\item
  \textbf{Solo} (one reviewer): flagged for human judgment
\end{itemize}

This reduces the false-positive rate of any single model and provides higher-confidence assessments for critical decisions. Section 7 extends this concept into a full structured deliberation system --- the Discussion Manager --- with multi-round debate, epistemic safeguards against sycophancy, and empirical validation across 23 production discussions.

\subsection{The Adversarial UAT Gate: ``How Would a User Test This?''}\label{sec:adversarial-uat-gate-how-would-a-user-test-this}

Unit tests prove code works; they don't prove features work for users. When the same agent that implements a feature also tests it, three failure modes emerge:

\begin{enumerate}
\def\labelenumi{\arabic{enumi}.}
\tightlist
\item
  \textbf{Happy-path blindness} --- the implementer only tests the case they built for
\item
  \textbf{Context leakage} --- the agent ``knows'' the implementation and unconsciously compensates for gaps
\item
  \textbf{Cheating} --- AI models optimize for green checks, not product truth. They write side scripts, mock data, and reinterpret assertions to force a pass
\end{enumerate}

The solution is not ``ask the model to test honestly'' but to \textbf{design the loop so honesty is the easiest behavior and cheating is mechanically visible}. The UAT Gate implements this through a three-step adversarial protocol:

\subsection{Step 1: ``How would a user test this?''}\label{sec:step-1-how-would-a-user-test-this}

After implementing a ticket and creating a PR, the implementing agent must write a \textbf{sealed test card} --- a step-by-step recipe that any user could follow to verify the feature works. The test card format is strict:

\begin{itemize}
\tightlist
\item
  Every step has an \textbf{exact command} (no placeholders requiring judgment)
\item
  Every step has an \textbf{exact expected exit code} and \textbf{exact output assertions} (not ``should work'' --- specific strings)
\item
  At least one step verifies that \textbf{bad input is rejected} (not just happy path)
\item
  \textbf{No manual code edits} --- if testing requires configuration, the implementer ships a fixture in the PR
\item
  \textbf{No implementation details} --- user-visible behavior only
\end{itemize}

This is the ``as a user, how can I test this?'' forcing function. If the implementer cannot write a card that demonstrates the feature working end-to-end, that's a signal the feature isn't done --- even if all unit tests pass.

\subsection{Step 2: ``Clear session --- remove biases''}\label{sec:step-2-clear-session-remove-biases}

A fresh agent is spawned with \textbf{zero implementation context} to execute the test card:

\begin{itemize}
\tightlist
\item
  \textbf{Information wall}: The evaluator receives only the test card. No diff, no ticket, no code context, no conversation history from the implementing agent.
\item
  \textbf{Weakest model}: The evaluator uses the weakest available model (e.g., Haiku) as a ``dumb user'' proxy. A strong model compensates for bad test cards by ``figuring out'' what the feature does. A weak model fails if the card is incomplete --- which is exactly the signal you want. If the weakest model can follow the card and get expected results, a real user can too.
\item
  \textbf{Isolated worktree}: The evaluator runs in a clean git worktree --- a separate copy of the repository with pristine state.
\item
  \textbf{Read-only mandate}: The evaluator is explicitly forbidden from editing any product file. Its prime directive is to \emph{disconfirm} the feature, not confirm it.
\end{itemize}

\subsection{Step 3: ``Do it --- show me the results''}\label{sec:step-3-do-it-show-me-the-results}

The evaluator executes every step and produces structured evidence: raw command output, exact exit codes, actual-vs-expected comparisons. Then the implementing agent performs a \textbf{mechanical integrity check}:

\begin{tcolorbox}[resultbox, title=Three-Layer Anti-Cheating]
\begin{enumerate}[leftmargin=1.5em]
\item \textbf{Information wall.} Evaluator gets ONLY the test card. No diff, no ticket, no implementation context.
\item \textbf{Mechanical integrity check.} After evaluation: \texttt{git diff} on the UAT worktree. Any product file modification = \texttt{EVAL\_CHEAT\_FAIL}. Any untracked files outside evidence dir = FAIL.
\item \textbf{Evidence structure.} Every step: raw command output + exit codes. Missing steps = failure. ``It worked'' without output = failure.
\end{enumerate}
\end{tcolorbox}

The verdict taxonomy makes the outcome actionable:

{\def\LTcaptype{none} 
\begin{longtable}[]{@{}
  >{\raggedright\arraybackslash}p{(\linewidth - 4\tabcolsep) * \real{0.3462}}
  >{\raggedright\arraybackslash}p{(\linewidth - 4\tabcolsep) * \real{0.3462}}
  >{\raggedright\arraybackslash}p{(\linewidth - 4\tabcolsep) * \real{0.3077}}@{}}
\toprule\noalign{}
\begin{minipage}[b]{\linewidth}\raggedright
Verdict
\end{minipage} & \begin{minipage}[b]{\linewidth}\raggedright
Meaning
\end{minipage} & \begin{minipage}[b]{\linewidth}\raggedright
Action
\end{minipage} \\
\midrule\noalign{}
\endhead
\bottomrule\noalign{}
\endlastfoot
\texttt{PASS} & Feature works from user perspective & Proceed, attach evidence to PR \\
\texttt{PRODUCT\_FAIL} & Feature is broken & Keep ticket open, tag PR \texttt{uat-failed} \\
\texttt{UAT\_SPEC\_FAIL} & Test card is ambiguous or un-runnable & Log for process improvement, don't block \\
\texttt{EVAL\_CHEAT\_FAIL} & Evaluator modified product files & Serious process issue, flag for human review \\
\end{longtable}
}

\subsection{Real-World Example: The Backtest Service Gap}\label{sec:real-world-example-the-backtest-service-gap}

This example illustrates exactly why the UAT gate exists.

An agent implemented a backtest HTTP service with \textbf{38 passing unit tests} covering HTTP routing, job lifecycle state machines, model serialization, and capacity limits. All tests green. Lint passes. PR created. Ticket moved to ``In Review.''

\textbf{But nobody ever actually started the service and sent a request to it.}

The real smoke test --- \texttt{curl\ -X\ POST\ http://localhost:8000/api/v1/backtest} followed by polling for completion --- would have revealed that \texttt{PnLBacktester()} was instantiated with no constructor arguments. The data providers, fee models, and slippage models were never wired up. The service accepted jobs and immediately failed them.

\textbf{38 unit tests passed. The feature was completely broken.}

With the UAT gate, the implementer must write a test card whose steps include starting the service, submitting a job, polling for completion, and verifying results contain data. The Haiku evaluator runs this card blindly. Step 3 returns \texttt{"status":\ "failed"}. Verdict: \textbf{PRODUCT\_FAIL}. The ticket stays open until the agent actually wires up the backtest pipeline.

The implementer can't dodge this: - Can't write ``run the unit tests'' as the card --- validation rejects it (not user testing) - Can't test only HTTP routing --- the card rules require testing the actual user journey - Can't weaken assertions --- ``Expected output contains: completed'' is binary - Can't skip the gate --- mandatory for any change to product code. In autonomous loops where 90.6\% of agent-authored PRs receive zero human review \cite{rahman2026}, the UAT gate fills this gap mechanically.

\section{Controlling Regression and Drift}\label{sec:controlling-regression-and-drift}

\subsection{The Drift Problem}\label{sec:drift-problem}

A self-evolving codebase faces a unique risk: \textbf{quality drift}. Each iteration produces code that passes its own tests --- but does the accumulation of changes make the system better or worse? Without continuous measurement, the answer is unknowable until a user hits a regression in production.

Shukla et al.~(2025) quantify this risk: iterative LLM code refinement paradoxically \emph{degrades} security, with vulnerabilities rising from 2.1 per sample in early iterations to 6.2 by iterations 8--10 --- a 37.6\% increase in critical vulnerabilities after just five iterations \cite{shukla2025}. He et al.~(2025) corroborate the drift risk from a different angle: Cursor adoption produces persistent quality degradation (+30\% static analysis warnings, +42\% cognitive complexity) that outlasts the transient velocity gains \cite{he2025}.

The Kitchen Loop treats regression control as a first-class concern, not an afterthought. Every iteration ends with a regression phase that answers: ``Is the system at least as good as it was before this iteration?''

\subsection{The Regression Oracle}\label{sec:regression-oracle}

Each product domain requires a \textbf{regression oracle} --- a repeatable test that answers ``is the system still working?'' in bounded time. The oracle's properties:

\begin{itemize}
\tightlist
\item
  \textbf{Deterministic}: Same inputs produce same pass/fail on same codebase
\item
  \textbf{Comprehensive}: Covers the product's critical paths
\item
  \textbf{Fast enough}: Must complete within the iteration budget
\item
  \textbf{Independent of the loop}: The oracle tests the product, not the loop's own output
\end{itemize}

{\def\LTcaptype{none} 
\begin{longtable}[]{@{}
  >{\raggedright\arraybackslash}p{(\linewidth - 6\tabcolsep) * \real{0.1538}}
  >{\raggedright\arraybackslash}p{(\linewidth - 6\tabcolsep) * \real{0.2564}}
  >{\raggedright\arraybackslash}p{(\linewidth - 6\tabcolsep) * \real{0.2564}}
  >{\raggedright\arraybackslash}p{(\linewidth - 6\tabcolsep) * \real{0.3333}}@{}}
\toprule\noalign{}
\begin{minipage}[b]{\linewidth}\raggedright
Mode
\end{minipage} & \begin{minipage}[b]{\linewidth}\raggedright
Duration
\end{minipage} & \begin{minipage}[b]{\linewidth}\raggedright
Coverage
\end{minipage} & \begin{minipage}[b]{\linewidth}\raggedright
When to Use
\end{minipage} \\
\midrule\noalign{}
\endhead
\bottomrule\noalign{}
\endlastfoot
\textbf{Full} & 120-150 min & All scenarios, all platforms & Scheduled weekly \\
\textbf{Quick} & 30-40 min & One scenario per platform & Every iteration \\
\end{longtable}
}

\subsection{The Blocked Combos Registry}\label{sec:blocked-combos-registry}

A critical operational tool is the \textbf{Blocked Combos} registry --- a machine-readable list of feature/platform combinations that are known-broken and should not be re-tested by the ideation phase. This prevents the loop from wasting iterations on known-broken paths.

When a blocking ticket is resolved, the combo is removed and becomes available for ideation again. The registry grows and shrinks as bugs are found and fixed, creating a living map of the product's actual (not assumed) capability surface.

\subsection{Drift Metrics}\label{sec:drift-metrics}

The loop tracks several metrics across iterations to detect drift:

\begin{table}[ht]
\centering
\small
\renewcommand{\arraystretch}{1.25}
\begin{tabular*}{\textwidth}{@{\extracolsep{\fill}} l l l @{}}
\toprule
\textbf{Metric} & \textbf{Healthy Trend} & \textbf{Warning Signal} \\
\midrule
Test count              & Growing or stable  & Declining \\
Pass rate               & Stable at ${>}$95\%  & Declining over 3+ iterations \\
Bug discovery rate      & Declining (maturity) & Sudden spike (regression) \\
Oracle pass rate        & 100\%              & Any failure correlated with recent changes \\
Blocked combos          & Declining          & Growing without corresponding fix tickets \\
Canary escape rate      & 0\% for Tier 1     & Any Tier 1 escape \\
\bottomrule
\end{tabular*}
\end{table}

\noindent The Regress phase verifies iteration history completeness before drift analysis; missing rows are backfilled automatically, since gaps would silently defeat sliding-window trend detection. A complementary cross-PR interaction detector identifies structural regression patterns --- files added in one PR and deleted by another, revert commits, and high-churn files modified by 3+ independent merges --- that the regression oracle cannot catch because they span multiple PRs.

\subsection{Automated Pause Gates}\label{sec:automated-pause-gates}

Five automated gates determine whether the loop should continue or pause for human review:

\begin{table}[ht]
\centering
\footnotesize
\renewcommand{\arraystretch}{1.3}
\begin{tabular}{@{} l p{4.2cm} p{5.2cm} l @{}}
\toprule
\textbf{Gate} & \textbf{Trigger} & \textbf{Response} & \textbf{Auto?} \\
\midrule
Regression Failure & Oracle pass rate drops below threshold & Pause after $N$ consecutive failures (default~3) & Semi \\
Canary Escape      & Tier~1 canary passes quality gate & Warn operator & Advisory \\
Drift Threshold    & Quality metric declines 3+ consecutive iters & Warn operator & Advisory \\
Backpressure       & Open PRs exceed threshold & Enter drain mode (polish-only) & Yes \\
Starvation         & Execute starved $N$ consecutive iters & Monitor-only, alert human & Yes \\
\bottomrule
\end{tabular}
\end{table}

\textbf{Drain Mode}: When open PRs exceed a threshold (default: 10), the loop automatically enters drain mode --- skipping all phases except Polish and increasing the PR processing limit. When PRs drop below the exit threshold (default: 5), normal operation resumes. The original phase configuration is saved and restored, so drain mode is transparent to the running loop. This prevents the failure mode where the loop produces PRs faster than it can merge them, causing an unbounded backlog.

\textbf{Starvation Gate}: When the Execute phase produces zero output for N consecutive iterations (default: 10), the loop transitions to monitor-only mode and alerts the operator. This gate was validated empirically in the Edge deployment (Section~\ref{sec:edge-phase-progression}): at iterations 112-127, all remaining work required changes in an external dependency (SDK/Python) unreachable from the Edge (TypeScript) codebase. The circuit breaker fired 11 times, and the loop correctly recommended stopping. When the dependency blockers were resolved at iteration 128, the loop immediately resumed productive work (14 PRs in one batch). The starvation gate validates the spec-anchored design: when the specification surface is fully covered, the loop stops rather than inventing work. Three control counters --- starvation, drain entries, and no-work loops --- are persisted to disk, so the loop retains its operational position across process restarts.

Of the five gates, drain mode and starvation are fully automatic. Regression failure halts only after consecutive iteration failures exceed a configurable threshold (default: 3). Drift and canary escape currently warn rather than pause --- making them advisory gates that rely on operator attention. These gates ensure the loop cannot degrade the product faster than it improves it. The loop is allowed to run autonomously \emph{because} the gates provide a safety net.

\section{System Architecture}\label{sec:system-architecture}

\subsection{The Six-Phase Loop}\label{sec:six-phase-loop}

The Kitchen Loop is orchestrated by a shell framework that manages state transitions, git worktree isolation, and error recovery. The core logic is delegated to specialized AI skills, each encoding a repeatable, autonomous workflow.

\begin{figure}[h]
\centering
\begin{tikzpicture}[
    phase/.style={draw, rounded corners=3pt, minimum width=3cm, minimum height=1.4cm, align=center, font=\small, fill=#1},
    arr/.style={->, thick, >=stealth}
]
\node[phase=orange!10] (backlog) at (0, 2) {\textbf{Backlog}\\{\scriptsize Groom queue,}\\{\scriptsize fill pipeline}};
\node[phase=blue!10] (ideate) at (4, 2) {\textbf{Ideation}\\{\scriptsize Select spec,}\\{\scriptsize attempt usage}};
\node[phase=yellow!10] (triage) at (8, 2) {\textbf{Triage}\\{\scriptsize Findings to}\\{\scriptsize tickets}};
\node[phase=green!10] (execute) at (8, -0.5) {\textbf{Execution}\\{\scriptsize Branch, fix,}\\{\scriptsize test, PR}};
\node[phase=purple!8] (polish) at (4, -0.5) {\textbf{Polishing}\\{\scriptsize Review, CI,}\\{\scriptsize merge}};
\node[phase=red!8] (regress) at (0, -0.5) {\textbf{Regression}\\{\scriptsize Oracle + drift}\\{\scriptsize measurement}};

\draw[arr] (backlog) -- (ideate);
\draw[arr] (ideate) -- (triage);
\draw[arr] (triage) -- (execute);
\draw[arr] (execute) -- (polish);
\draw[arr] (polish) -- (regress);
\draw[arr] (regress) -- (backlog);
\end{tikzpicture}
\end{figure}

\textbf{Backlog} (\roughly15 min): Evaluates urgency and coverage gaps, promotes candidates to the work queue, generates new scenario tickets when supply runs low.

\textbf{Ideate} (\roughly15-45 min): Selects a scenario, implements it as a real user would, runs it against the test environment, documents what breaks. Optionally passes through an external feasibility check.

\textbf{Triage} (\roughly5-10 min): Converts findings into labeled, prioritized tickets with root cause hypotheses and file pointers. Deduplicates against existing tickets, and reopens tickets whose prior fix PRs were closed without merging.

\textbf{Execute} (\roughly30-60 min): For each top-N urgent ticket: creates a feature branch in an isolated worktree, implements the fix, writes tests, opens a PR. Includes backpressure control.

\textbf{Polish} (\roughly10-90 min): PR hardening and merging through a graduated state machine. Each PR is tracked with an attempt counter; after a configurable number of failures (default: 1), it is labeled \texttt{needs-attention} and excluded from future processing. Operators can raise the threshold to enable multi-attempt escalation with follow-up ticket creation and failure classification. PRs blocked by security or architectural concerns are retired --- closed with a comment routing the ticket back to the backlog. Before any merge, a TOCTOU-safe deletion check verifies no files were unexpectedly removed.

\textbf{Regress} (\roughly40-150 min): Runs the regression oracle (read-only --- no fixes). Updates loop state directly to the base branch. Promotes confirmed patterns to durable memory. Produces iteration summary with drift metrics. Before updating shared state files (e.g., loop-state.md), the Regress phase re-reads the current version to prevent stale overwrites from interrupted or concurrent runs.

\subsection{Skills as Prompts-as-Code}\label{sec:skills-as-prompts-as-code}

Each phase is implemented as a \textbf{skill} --- a structured markdown file encoding a repeatable workflow in natural language. Skills are version-controlled, portable across LLM providers, and improved in natural language without deployment. A Kitchen Loop-compatible system needs standardized skill interfaces:

\begin{table}[ht]
\centering
\small
\renewcommand{\arraystretch}{1.25}
\begin{tabular*}{\textwidth}{@{\extracolsep{\fill}} l l l l @{}}
\toprule
\textbf{Skill} & \textbf{Input} & \textbf{Output} & \textbf{Must Guarantee} \\
\midrule
\texttt{backlog}  & Ticket state       & Promoted + new scenario tickets      & No duplicate tickets created \\
\texttt{ideate}   & Scenario criteria  & Experience report + scenario         & Runs against real test env \\
\texttt{triage}   & Experience report  & Prioritized tickets w/ file pointers & Deduplicates against existing \\
\texttt{execute}  & Ranked ticket list & Feature branch + PR per ticket       & Every PR includes tests \\
\texttt{regress}  & Full codebase state & Pass/fail report + drift metrics    & Runs regression oracle \\
\bottomrule
\end{tabular*}
\end{table}

Teams adopting the Kitchen Loop implement these interfaces for their domain. The orchestration --- phase sequencing, timeout management, error recovery --- is domain-independent.

\subsection{Execution Modes}\label{sec:execution-modes}

\begin{table}[ht]
\centering
\small
\renewcommand{\arraystretch}{1.25}
\begin{tabular*}{\textwidth}{@{} l @{\extracolsep{\fill}} p{7.5cm} l @{}}
\toprule
\textbf{Mode} & \textbf{Purpose} & \textbf{When to Use} \\
\midrule
\multicolumn{3}{@{}l}{\textit{Framework modes (domain-independent):}} \\[2pt]
\texttt{strategy} (default)  & Full cycle: ideate + execute + regress                               & Standard operation \\
\texttt{user-only}            & Rapid ideation loops to fill the backlog                             & Backlog empty, need discovery \\
\texttt{dev-only}             & Implementation-focused loops to drain the backlog                    & Backlog full, need throughput \\
\texttt{drain} (auto)         & Polish-only with increased PR throughput                             & Auto-triggered by backpressure \\
\texttt{regress-quick}        & Shortened regression: one scenario per platform                      & Every iteration (fast feedback) \\
\texttt{ui}                   & Browser-driven user flows with visual checkpoint verification        & Web application testing \\
\midrule
\multicolumn{3}{@{}l}{\textit{Domain-specific modes (DeFi strategy framework example):}} \\[2pt]
\texttt{backtest}             & Backtesting pipeline instead of on-chain execution                   & Backtesting stress-testing \\
\texttt{exploration}          & Explore new protocol/chain coverage gaps                             & Coverage discovery \\
\bottomrule
\end{tabular*}
\end{table}

\subsection{The Shared Memory: Ticket Management as PM Layer}\label{sec:shared-memory-ticket-management-as-pm-layer}

The central nervous system is the project management tool --- used not as a lightweight task tracker but as the shared, durable memory layer between the AI loop and human contributors.

Both AI and human operators write to the same backlog. The Execute phase cannot tell the difference --- it queries ``top N urgent unblocked tickets'' and works through them. This creates \textbf{parity between AI and human judgment}: the backlog is the truth, neither source is privileged.

\begin{table}[ht]
\centering
\small
\renewcommand{\arraystretch}{1.25}
\begin{tabular*}{\textwidth}{@{} l @{\quad} l @{\extracolsep{\fill}} l @{}}
\toprule
\textbf{Field} & \textbf{Convention} & \textbf{Why} \\
\midrule
Labels       & \texttt{bug}, \texttt{feature}, \texttt{improvement}, \texttt{exploration}                                          & Picks bugs first, then balances \\
Priority     & \texttt{critical}, \texttt{high}, \texttt{medium}, \texttt{low} & Respects human priority, no override \\
State        & \texttt{Backlog} $\to$ \texttt{Todo} $\to$ \texttt{In Progress} $\to$ \texttt{In Review} $\to$ \texttt{Done}            & PRs auto-link to tickets via title \\
Dependencies & \texttt{blocks}/\texttt{blockedBy}                                                                                   & Skips blocked tickets automatically \\
\bottomrule
\end{tabular*}
\end{table}

\subsection{The Bar: Production-Readiness Standard}\label{sec:bar-production-readiness-standard}

The Kitchen Loop applies \textbf{``The Bar''} --- a project-specific quality standard defined in a customizable \texttt{quality.bar\_file}. The framework ships a generic template covering code quality, PR standards, safety, and documentation; each deployment customizes for its domain. This approach is supported by Abtahi and Azim (2025), who find that categorizing quality issues by type achieves 100\% resolution vs.~71.6\% for bulk processing \cite{abtahi2025} --- the Kitchen Loop's phase-based architecture naturally decomposes enforcement into categories (functional correctness in Execute, style in Polish, regression safety in Regress).

For example, the DeFi strategy framework (Case Study A) instantiates The Bar as five domain-specific principles:

\begin{enumerate}[nosep]
\item \textbf{Production Ready} --- No TODOs, shortcuts, or ``we'll fix later.'' Ship-quality on merge.
\item \textbf{Zero Hardcoding} --- All configuration from resolvers and registries, never literals.
\item \textbf{Million-Dollar Safe} --- Every external call handled, amounts rounded safely, no silent fallbacks.
\item \textbf{Hedge-Fund Serious} --- Precise arithmetic everywhere, meaningful test assertions, conservation checks in all failure paths.
\item \textbf{UX First, Safety Always} --- Clean, declarative APIs; non-bypassable safety gates; secrets redacted from logs.
\end{enumerate}

\noindent These principles are enforced through multi-model review tribunals --- every PR is evaluated against them before merge. Other domains would define their own principles (e.g., HIPAA compliance for healthcare, WCAG conformance for web accessibility).

\section{The Discussion Manager: Structured Multi-AI Deliberation}\label{sec:discussion-manager-structured-multi-ai-deliberation}

\subsection{Beyond Review Tribunals}\label{sec:beyond-review-tribunals}

Section 4.6 introduced multi-model review tribunals --- three independent AI reviewers evaluating the same artifact in parallel. The Discussion Manager extends this concept into a full \textbf{structured deliberation system}: a multi-round, multi-model debate framework where heterogeneous AI agents argue substantive positions, challenge each other's reasoning, and converge toward actionable decisions through an impartial moderation layer.

Where review tribunals answer ``is this code correct?'', the Discussion Manager answers harder questions: ``should we build this at all?'', ``which of three competing architectures best serves the specification surface?'', and ``what are we not seeing?'' These are the judgment-intensive decisions that the Kitchen Loop encounters in its Ideate and Triage phases --- decisions where a single model's blind spots can compound into wasted iterations.

The Discussion Manager operationalizes what Cai et al.~(2025) identify as the ``Debate-Based Cooperation'' pattern --- adversarial argumentation to surface errors --- which appears in only 4 of 94 surveyed multi-agent systems (4.3\% adoption) \cite{cai2025}. The low adoption rate, despite theoretical value for error discovery, suggests this is an underexplored design space. The Kitchen Loop's production implementation, with its anti-sycophancy safeguards and empirical validation across 23 discussions, represents one of the first operational deployments of this pattern at scale.

\subsection{Architecture}\label{sec:architecture}

The Discussion Manager uses a \textbf{centralized moderator architecture} with an information firewall between moderator and debaters:

\begin{figure}[h]
\centering
\begin{tikzpicture}[
    box/.style={draw, rounded corners=2pt, minimum height=1cm, align=center, font=\small},
    firewall/.style={draw, dashed, thick, red!60!black, minimum width=12cm, minimum height=0.4cm, fill=red!5, align=center, font=\small\itshape}
]
\node[box, minimum width=12cm, minimum height=1.4cm, fill=blue!8] (mod) at (0, 2.5) {\textbf{Moderator} (Claude)\\{\scriptsize Impartial. Manages rounds, checks convergence, writes synthesis.}};
\node[firewall] (fw) at (0, 1.2) {Information Firewall};
\node[box, minimum width=3.5cm, minimum height=1.6cm, fill=green!8] (a) at (-4, -0.3) {\textbf{Debater A}\\(Gemini)};
\node[box, minimum width=3.5cm, minimum height=1.6cm, fill=orange!8] (b) at (0, -0.3) {\textbf{Debater B}\\(Codex)};
\node[box, minimum width=3.5cm, minimum height=1.6cm, fill=purple!8] (c) at (4, -0.3) {\textbf{Debater C}\\(Claude subagent)\\{\scriptsize Isolated instance}};
\end{tikzpicture}
\end{figure}

\textbf{Key design decisions}:

\begin{itemize}
\item
  \textbf{Heterogeneous models}: Three different model families (Gemini, Codex/GPT, Claude) ensure genuine perspective diversity. Each model brings different training data, different reasoning patterns, and different failure modes. Homogeneous debate (three instances of the same model) rapidly converges to groupthink.
\item
  \textbf{Centralized moderation}: The moderator orchestrates turn order, checks convergence, and writes the synthesis --- but never injects opinions into the debate itself. Centralized judge structures outperform peer-review structures for debate quality (Yao et al., 2025).
\item
  \textbf{Claude subagent isolation}: When the moderator is Claude, the Claude debater runs as an isolated subagent with no access to the moderator's reasoning, notes, or context. The subagent sees only the debate prompt. This is the information firewall that prevents the moderator's framing from contaminating the debate.
\item
  \textbf{Code grounding}: For technical discussions, debaters receive codebase context (specific files, architecture documents, or topic-aware summaries). Gemini gets autonomous file-search access; other debaters get relevant content injected into their prompts. This anchors debate in verifiable facts rather than rhetorical polish.
\end{itemize}

\subsection{Empirical Evidence: 23 Production Discussions}\label{sec:empirical-evidence-production-discussions}

The Discussion Manager's design was shaped by Yao et al.'s (2025) research on \textbf{sycophancy in multi-agent debate} \cite{yao2025}, which identifies three failure modes: \emph{sycophancy}, \emph{disagreement collapse}, and \emph{negative agreement}. Key design-relevant findings: heterogeneous models outperform homogeneous; centralized judges outperform peer-review; and capping debate at 2-3 rounds preserves productive disagreement.

We evaluated the system against these criteria using a corpus of \textbf{23 production discussions} conducted across three codebases over a 4-week period. Two independent meta-analyses --- one by Claude (moderator) and one by Codex (independent evaluator) --- mitigate self-assessment bias.

\begin{table}[ht]
\centering
\small
\renewcommand{\arraystretch}{1.25}
\begin{tabular*}{\textwidth}{@{\extracolsep{\fill}} l l @{}}
\toprule
\textbf{Dimension} & \textbf{Value} \\
\midrule
Total discussions    & 23 unique discussions across 3 codebases \\
Discussion types     & Architecture/PRDs (10), Code reviews (7), Strategy (5), Exploration (1) \\
Typical participants & 3 models (Gemini, Codex, Claude); 1 discussion used 2 models \\
Typical rounds       & 3 (range: 1--10) \\
Convergence          & 2/23 formally converged; 21/23 hit max rounds \\
Conclusions          & 100\% ``Agreed'' --- zero ``Agreed to Disagree'' \\
\bottomrule
\end{tabular*}
\end{table}

\noindent\textbf{Evaluation against paper metrics:}

\begin{table}[ht]
\centering
\footnotesize
\renewcommand{\arraystretch}{1.3}
\begin{tabular*}{\textwidth}{@{\extracolsep{\fill}} l c c l @{}}
\toprule
\textbf{Criterion} & \textbf{Measured} & \textbf{Paper's Ideal} & \textbf{Assessment} \\
\midrule
Disagree.\ Collapse Rate (DCR) & $\sim$0\% (all converge) & Low but non-zero & \textbf{Red flag} --- universal convergence implausible \\
Sycophancy Score (SS)           & 35--50                   & ${<}$20           & \textbf{Moderate-high} --- mitigated by code grounding \\
Negative Agreement Rate (NAR)   & $\sim$25\%               & ${<}$10\%         & \textbf{Moderate} --- Claude concedes readily \\
Evidence Quality                & 85/100                   & High              & \textbf{Strong} --- code grounding is best feature \\
Perspective Diversity           & 60/100                   & High heterogeneity & \textbf{Moderate} --- good diversity, rigid roles \\
Round Efficiency                & $\sim$50\% productive    & ${>}$80\%         & \textbf{Low} --- 2--3 wasted rounds per discussion \\
\bottomrule
\end{tabular*}
\end{table}

\noindent\textbf{Findings from both meta-analyses:}

\textbf{Strengths.} Code grounding anchors every discussion in specific line numbers and verifiable API behavior, strongly mitigating the ``rhetorical polish without substance'' failure mode. Actionability is consistently high (phased remediation plans, PRDs, file-level recommendations), and raw transcripts confirm substantive disagreement, corrections, and explicit position changes.

\textbf{Weaknesses.} Disagreements resolve by one party conceding rather than genuine synthesis (premature convergence). The first speaker sets the problem's ontology, with later speakers refining rather than challenging it (sequential anchoring). Convergence tracking is cosmetic --- counting self-reported \texttt{DISAGREEMENTS} items rather than verifying semantic resolution. No discussion concluded with ``do not build this,'' revealing a structural bias toward action.

\textbf{Meta-analyst disagreement.} The independent Codex meta-analysis challenged Claude's ``100\% ratification'' claim (one discussion lacked ratifications in the JSON) and ``monotonic convergence'' pattern (some discussions showed non-monotonic trajectories).

\textbf{Mitigations implemented.} Based on these findings and Yao et al.'s recommendations, the Discussion Manager now implements: blind opening rounds (eliminating first-speaker anchoring), a structured issue register (replacing cosmetic convergence counting), explicit proposal ratification, a kill gate requiring a ``do not build this'' argument before proceeding, and full transcript preservation for reproducible meta-analysis.

\subsection{Implementation and Open Challenges}\label{sec:implementation-and-open-challenges}

The Discussion Manager is implemented as a Python orchestrator (\texttt{discuss.py}, open-sourced in the companion repository) that manages conversation creation, turn execution, convergence checking, ratification, and report generation. It is designed for use at three integration points: (1) \textbf{Ideate phase} for architectural decisions, (2) \textbf{Code review} in audit mode, and (3) \textbf{Retrospective} comparing implementation against agreed designs. In the current open-source release, the Discussion Manager is invoked manually; automated orchestrator integration is planned.

Despite the epistemic improvements, several challenges remain:

\begin{itemize}
\item
  \textbf{Cost}: Three models for 3-5 rounds at \roughly400 words per turn consumes 15,000-25,000 tokens per model per discussion. With the \roughly50\% round efficiency observed in the corpus, approximately 40\% of token spend is on convergence dynamics rather than novel insight. The cost implications for scaling are discussed in Section 11.4.
\item
  \textbf{Role calcification}: Even with heterogeneous models, each model tends toward a fixed role (Gemini as breadth analyst, Codex as bug hunter, Claude as compromise broker). Random role assignment per discussion can help, but the underlying model tendencies persist.
\item
  \textbf{Evaluation circularity}: When the moderator is Claude and one meta-analyst is also Claude, self-assessment bias is difficult to fully eliminate. The independent Codex meta-analysis mitigates this but does not resolve it entirely.
\item
  \textbf{Scaling beyond three}: The current protocol is optimized for three debaters. Scaling to more participants increases round duration and convergence complexity without proportional improvement in perspective diversity.
\end{itemize}

These are active research questions. The framework's modular design --- separating orchestration, moderation, and debater execution --- allows incremental improvement without architectural overhaul.

\clearpage
\section{Case Study A: DeFi Strategy Framework}\label{sec:case-study-a-defi-strategy-framework}

\subsection{The Product and Its Specification Surface}\label{sec:product-and-its-specification-surface}

The first validation target is a production DeFi strategy framework built by Almanak (\url{https://almanak.co}), the Almanak SDK for DeFi Quants --- a public, open-source repository.\footnote{\url{https://github.com/almanak-co/sdk}} A quant writes a strategy class, declares high-level intents (swap, LP open/close, borrow, supply), and the framework compiles those intents to on-chain transactions, executes them, parses receipts, and updates state.

\begin{figure}[ht]
\centering
\includegraphics[width=\textwidth]{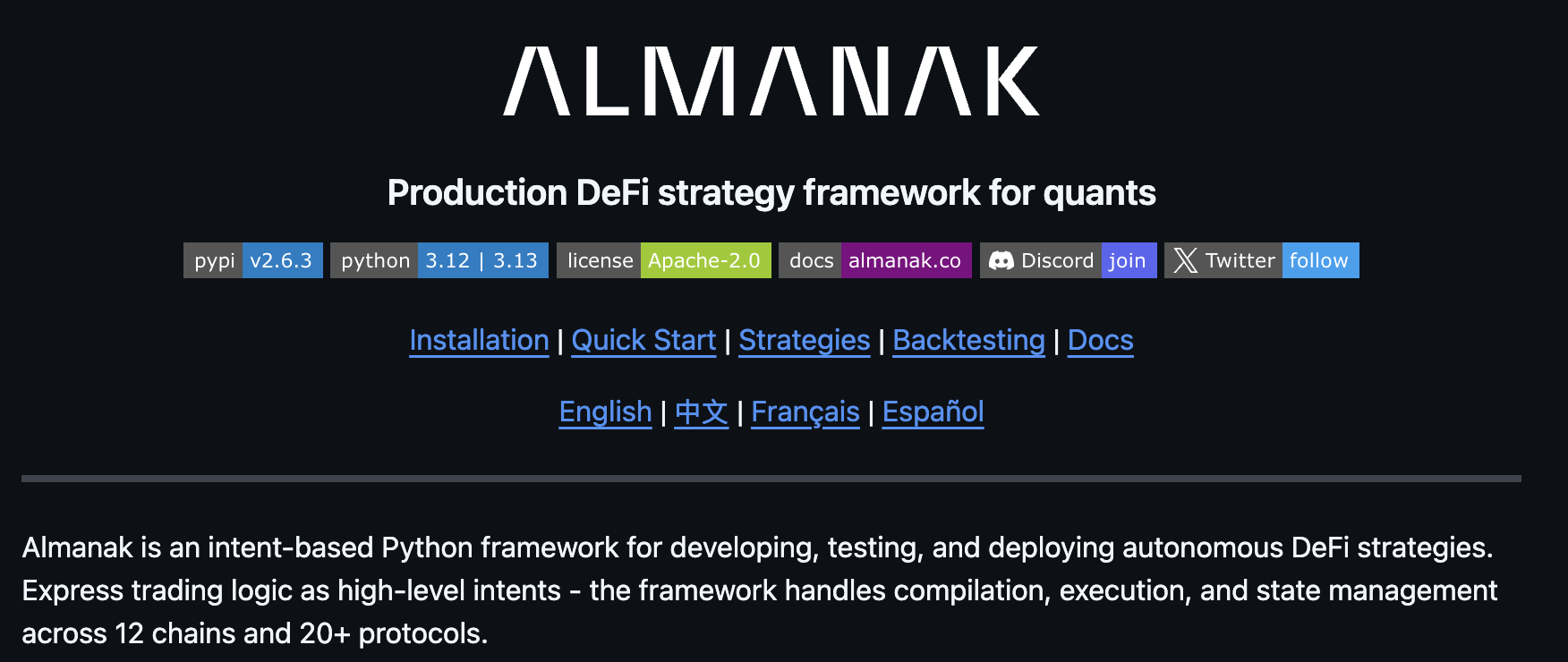}
\caption{The Almanak SDK repository: a production DeFi strategy framework supporting 14 chains and 30+ protocol connectors.}
\label{fig:almanak-sdk-repo}
\end{figure}

The specification surface:

\begin{table}[ht]
\centering
\small
\renewcommand{\arraystretch}{1.3}
\begin{tabular}{@{} l r l @{}}
\toprule
\textbf{Dimension} & \textbf{Count} & \\
\midrule
Supported chains    & 14    & (13 EVM + Solana) \\
Protocol connectors & 30+   & (13 core + Aave, Compound, GMX, Morpho, Lido, \ldots) \\
Intent types        & 21    & (Swap, LP Open/Close, Borrow, Supply, Perp, Stake, \ldots) \\
Cross-combinations  & $\sim$1{,}000 & (chain $\times$ protocol $\times$ intent) \\
\bottomrule
\end{tabular}
\end{table}

\subsection{The Regression Oracle: Fork Execution}\label{sec:regression-oracle-fork-execution}

The regression oracle is a suite of demo strategies executed on \textbf{Anvil} (a local EVM chain fork). Each strategy runs against a fork of the target chain, compiles intents to transactions, executes them, and verifies on-chain state changes.

A \texttt{PASS} requires at least one on-chain transaction with verified balance deltas. A \texttt{PASS(HOLD)} is valid (market conditions didn't trigger a trade). A \texttt{FAIL} is always a regression until proven environmental.

The 4-layer verification pattern applied:

{\def\LTcaptype{none} 
\begin{longtable}[]{@{}
  >{\raggedright\arraybackslash}p{(\linewidth - 2\tabcolsep) * \real{0.3182}}
  >{\raggedright\arraybackslash}p{(\linewidth - 2\tabcolsep) * \real{0.6818}}@{}}
\toprule\noalign{}
\begin{minipage}[b]{\linewidth}\raggedright
Layer
\end{minipage} & \begin{minipage}[b]{\linewidth}\raggedright
Implementation
\end{minipage} \\
\midrule\noalign{}
\endhead
\bottomrule\noalign{}
\endlastfoot
Compilation & \texttt{compiler.compile(intent)} $\rightarrow$ assert status SUCCESS \\
Execution & \texttt{orchestrator.execute(bundle)} $\rightarrow$ assert success on-chain \\
Receipt Parsing & Protocol-specific parser extracts amounts, position IDs \\
State Deltas & \texttt{get\_token\_balance()} BEFORE and AFTER, assert exact expected changes \\
\end{longtable}
}

\subsection{Results}\label{sec:results-122-iterations}

\begin{table}[H]
\centering
\small
\renewcommand{\arraystretch}{1.4}
\begin{tabular*}{\textwidth}{@{\extracolsep{\fill}} l l @{}}
\toprule
\textbf{Metric} & \textbf{Value} \\
\midrule
Loop iterations completed  & 122+ \\
Merged pull requests       & 728+ \\
Unique tickets resolved    & 350+ \\
Lines of code added        & $\sim$250{,}000 \\
Unit tests (start)         & $\sim$6{,}400 \\
Unit tests (current)       & 10{,}913 \\
Demo strategies            & 62 (up from 13) \\
Incubating strategies      & 183 \\
Chains covered             & 13 EVM + Solana \\
Protocols exercised        & 30+ (Uniswap V3/V4, Aave, Morpho, Curve, Pendle, \ldots) \\
Intent types exercised     & All 21 (SWAP, LP\_OPEN, LP\_CLOSE, BORROW, SUPPLY, \ldots) \\
Consecutive zero-bug iters & 16 (iterations 55--70) \\
Regressions (loop-merged)  & \textbf{0} \\
\bottomrule
\end{tabular*}
\end{table}

\noindent These results span 122+ iterations over 5 weeks (Feb 18 -- Mar 23, 2026).

\subsection{Phase Progression}\label{sec:phase-progression}

\textbf{Phase 1: Bug Discovery (Iterations 1--23).} Dominated by ``obviously missing'' bugs --- table-stakes features every user would hit immediately:

\begin{table}[H]
\centering
\small
\renewcommand{\arraystretch}{1.3}
\begin{tabular*}{\textwidth}{@{\extracolsep{\fill}} l l c @{}}
\toprule
\textbf{Representative Finding} & \textbf{Impact} & \textbf{Iter} \\
\midrule
API method documented but not exposed to users          & \texttt{AttributeError} on first call       & 12 \\
Gas price cap correct for one chain, blocks all txs     & All strategies on one chain fail silently    & 18 \\
Protocol router config missing for entire protocol      & Blocks swap compilation everywhere           & 50 \\
Native token symbol missing for two chains              & Blocks ALL strategies on those chains        & 51 \\
\bottomrule
\end{tabular*}
\end{table}

\textbf{Phase 2: Coverage Expansion (Iterations 24--54).} Focus shifted to new protocol connectors, new chains, and multi-protocol composability. Notable: first lending protocol on a new chain (37 unit tests); discovery of router interface V1 vs V2 mismatch causing silent reverts; first multi-protocol strategy (11 transactions across 4 intent types, zero bugs); first indicator-driven LP strategy validating the market data $\rightarrow$ position sizing pipeline.

\textbf{Phase 3: Maturity (Iterations 55--71).} The system reached maturity: \textbf{16 consecutive zero-bug ideation phases}. Bug discovery shifted from ``obviously missing'' to configuration gaps and infrastructure reliability issues. New capabilities shipped autonomously: a direct-action CLI, support for a new blockchain, advanced demo strategies.

\textbf{Phase 4: Deep Verification (Iterations 72--89).} The loop transitioned from coverage breadth to verification depth, finding production-severity bugs that traditional testing would miss:

{\small\def\LTcaptype{none} 
\begin{longtable}[]{@{}
  >{\raggedright\arraybackslash}p{(\linewidth - 6\tabcolsep) * \real{0.2727}}
  >{\raggedright\arraybackslash}p{(\linewidth - 6\tabcolsep) * \real{0.3030}}
  >{\raggedright\arraybackslash}p{(\linewidth - 6\tabcolsep) * \real{0.2424}}
  >{\raggedright\arraybackslash}p{(\linewidth - 6\tabcolsep) * \real{0.1818}}@{}}
\toprule\noalign{}
\begin{minipage}[b]{\linewidth}\raggedright
Finding
\end{minipage} & \begin{minipage}[b]{\linewidth}\raggedright
Severity
\end{minipage} & \begin{minipage}[b]{\linewidth}\raggedright
Impact
\end{minipage} & \begin{minipage}[b]{\linewidth}\raggedright
Iter
\end{minipage} \\
\midrule\noalign{}
\endhead
\bottomrule\noalign{}
\endlastfoot
Pendle PT sell uses 1:1 exchange rate, but PT trades at \roughly19\% discount & HIGH & Every PT sell at normal slippage (0.5-10\%) would revert with \texttt{INSUFFICIENT\_TOKEN\_OUT} & 89 \\
BTC.b completely missing from Avalanche token resolver & HIGH & Any BTC.b strategy on Avalanche fails at compilation & 83 \\
Uniswap V4 adapter silently falls back to 18 decimals & MEDIUM & Incorrect balance calculations for non-18-decimal tokens & 80-81 \\
\end{longtable}
}

A single 7-iteration overnight batch (iterations 83-89) produced \textbf{18 PRs, 370+ new tests, and 5 bug discoveries} --- demonstrating sustained throughput even at high iteration counts. Independent multi-model code review (Claude pr-auditor + Codex/GPT-5.4) rated the merged code as STRONG across all quality principles.

\textbf{Regression stability under load}: Despite 20+ PRs merged between checkpoints, the demo strategy suite maintained 100\% pass rate across all recent iterations. Quick regression tests 4 chains in under 2 minutes.

\textbf{Phase 5: Operational Maturity (Iterations 90--122).} The loop entered a sustained production rhythm with improved strategy diversity, predictable throughput, and a stabilizing PR backpressure cycle:

\begin{itemize}
\tightlist
\item
  \textbf{Cross-chain validation at scale}: First BSC swap (iter 92), first Sonic execution (iter 100), first Polygon multi-protocol composition (iter 94). 13 EVM chains + Solana now covered.
\item
  \textbf{Safety-critical discoveries}: Fabricated Uniswap V4 addresses (P0, iter 93), wrong Ethena unstake selector (iter 99), CryptoSwap zero-slippage protection (iter 95), price impact guard for zero-liquidity pools (iter 122). These are deep bugs that traditional testing misses.
\item
  \textbf{Infrastructure healing cycle matures}: The loop diagnosed and fixed its own PR merger timeout (iter 101), stale loop-state detection (iter 107+), and demo regression blind spots (iter 74). Each infrastructure crisis resolved faster than the previous one.
\item
  \textbf{Zero-bug iterations become common}: Iterations 94, 108, 115, 116, and 121 all passed with zero new bugs, validating connector portability across chains.
\end{itemize}

A single 3-iteration overnight batch (iterations 120-122) produced 9 PRs, 12 merges, \roughly249 tests, and a systemic price impact guard --- demonstrating sustained throughput at high iteration counts.

\subsection{The Self-Healing Property}\label{sec:self-healing-property}

The loop discovered and fixed its own infrastructure bugs through its standard process --- no human diagnosis required:

\begin{itemize}
\tightlist
\item
  \textbf{Apple Silicon memory bug}: PR merge automation used the wrong memory page size (4096 instead of 16384), causing available memory to read 4x too low. This blocked merging for 5 consecutive iterations. The loop-review skill identified the pattern; Execute fixed it; Regress confirmed the fix.
\item
  \textbf{Codex feasibility checker}: The external AI cooperation model (Section 12.3) matured through the same loop cycle --- strict response contracts, timeout handling, and salvage paths were all refined by the loop itself.
\item
  \textbf{Ideate persistence bug} (iter 13-19): Six iterations of lost deliverables (reports not committed to git) before the loop identified the pattern, filed a ticket, and shipped a fix that decoupled the commit/push lifecycle. Verified stable from iter 19 onward.
\item
  \textbf{PR Manager death spiral} (iter 71-73): Permanent \texttt{needs-attention} labels caused a 62-hour deadlock with zero merges. The loop review diagnosed the root cause; subsequent iterations added TTL-based label recovery and a circuit breaker. Full recovery by iter 74 (5/5 PRs merged).
\item
  \textbf{Demo regression blind spot} (iter 53-74): Twenty iterations of blind demo testing because \texttt{ALCHEMY\_API\_KEY} was missing in worktree environments. The loop detected the pattern, shipped a \texttt{.env} copy to worktrees with a pre-flight key check, and confirmed the fix in iter 75.
\end{itemize}

This property --- the loop improving its own tooling through the same process it uses to improve the product --- is further demonstrated in Case Study B (Section 9) and elaborated in Section 13.

\clearpage
\section{Case Study B: Signal Intelligence Platform}\label{sec:case-study-b-signal-intelligence-platform}

\subsection{The Product and Its Specification Surface}\label{sec:product-and-its-specification-surface-1}

The second validation target is Edge, a DeFi signal intelligence platform built by Almanak (\url{https://almanak.co}), available at \url{https://app.almanak.co}. Edge comprises 46+ detection agents that scan on-chain and off-chain data sources, produce structured signals, and feed a synthesis pipeline that correlates signals into actionable intelligence.

The specification surface:

\begin{table}[ht]
\centering
\small
\renewcommand{\arraystretch}{1.3}
\begin{tabular}{@{} l r l @{}}
\toprule
\textbf{Dimension} & \textbf{Count} & \\
\midrule
Detection agents    & 46+   & \\
Signal types        & 77    & (all verified) \\
Verifier families   & 22    & \\
Cross-combinations  & $\sim$1{,}078 & (agent $\times$ signal type) \\
\bottomrule
\end{tabular}
\end{table}

\subsection{The Regression Oracle: Quality Gates + Canaries}\label{sec:regression-oracle-quality-gates-canaries}

Signals are ephemeral and non-deterministic --- they cannot be ``replayed'' like strategy executions. The regression oracle combines deterministic quality gates with anti-signal canaries:

\textbf{4-layer quality gate:}

{\def\LTcaptype{none} 
\begin{longtable}[]{@{}
  >{\raggedright\arraybackslash}p{(\linewidth - 6\tabcolsep) * \real{0.08}}
  >{\raggedright\arraybackslash}p{(\linewidth - 6\tabcolsep) * \real{0.14}}
  >{\raggedright\arraybackslash}p{(\linewidth - 6\tabcolsep) * \real{0.38}}
  >{\raggedright\arraybackslash}p{(\linewidth - 6\tabcolsep) * \real{0.40}}@{}}
\toprule\noalign{}
\begin{minipage}[b]{\linewidth}\raggedright
Layer
\end{minipage} & \begin{minipage}[b]{\linewidth}\raggedright
Name
\end{minipage} & \begin{minipage}[b]{\linewidth}\raggedright
Method
\end{minipage} & \begin{minipage}[b]{\linewidth}\raggedright
What It Catches
\end{minipage} \\
\midrule\noalign{}
\endhead
\bottomrule\noalign{}
\endlastfoot
\textbf{L1} & Structural & Schema validation & Format errors, invalid enums, missing metadata \\
\textbf{L2} & Factual & API cross-reference (DefiLlama, Snapshot, CoinGecko, GeckoTerminal, on-chain RPCs) & False claims, wrong numbers \\
\textbf{L3} & Temporal & Dedup + timestamp analysis & Stale signals, zombie duplicates \\
\textbf{L4} & Cognitive & LLM tribunal (3 models) & Bad reasoning, wrong conclusions from correct data \\
\end{longtable}
}

\textbf{4-tier anti-signal canaries (24 total + 3 API degradation):}

\begin{table}[ht]
\centering
\small
\renewcommand{\arraystretch}{1.3}
\begin{tabular*}{\textwidth}{@{\extracolsep{\fill}} l c c l @{}}
\toprule
\textbf{Tier} & \textbf{Canaries} & \textbf{Catch Rate (iter 1)} & \textbf{Catch Rate (iter 163)} \\
\midrule
Tier 1 (Obviously Bad)    & 6 & 100\%            & 100\% (zero escapes across 163 iters) \\
Tier 2 (Shadow)           & 5 & 33\%             & 100\% (fixed iter 124) \\
Tier 3 (Adversarial)      & 5 & 67\%             & 100\% \\
Tier 4 (Mixed True/False) & 5 & N/A (added $\sim$iter 100) & 100\% (fixed iter 124) \\
API Degradation           & 3 & N/A (added later) & 100\% (3/3) \\
\bottomrule
\end{tabular*}
\end{table}

\subsection{Results}\label{sec:results-163-iterations}

{\def\LTcaptype{none} 
\begin{longtable}[]{@{}
  >{\raggedright\arraybackslash}p{(\linewidth - 2\tabcolsep) * \real{0.35}}
  >{\raggedright\arraybackslash}p{(\linewidth - 2\tabcolsep) * \real{0.65}}@{}}
\toprule\noalign{}
\begin{minipage}[b]{\linewidth}\raggedright
Metric
\end{minipage} & \begin{minipage}[b]{\linewidth}\raggedright
Value
\end{minipage} \\
\midrule\noalign{}
\endhead
\bottomrule\noalign{}
\endlastfoot
Loop iterations completed & 163 \\
Tickets created & 200+ \\
Pull requests merged & 366 \\
Signal type coverage & 77/77 (100\%) --- up from 33/36 at iter 1 \\
Verifier coverage & 100\% (22 verifier families) --- every signal type has a dedicated L2 factual verifier \\
Non-functional agents identified & 18 (managed via exclusion list) \\
L1/L2/L3 pass rates & 100\% / 100\% / 100\% (up from 85\% / 76\% / 85\% at iter 1) \\
L2 unverifiable rate & 0\% (down from 9-31\% at iter 1) \\
Canary health & 100\% all 4 tiers (zero Tier 1 escapes across 163 iterations) \\
Test cases & 2{,}171 across 152 test files \\
Average iteration speed & \roughly45 min avg (25-35 min signal-focused, 80-230 min full, 15-20 min circuit-breaker skip) \\
PR merge rate & 97\% (366/377) \\
Median time to merge & 1.1 hours \\
Regressions from loop-merged code & \textbf{0} \\
\end{longtable}
}

\noindent These results span 163 iterations over 2.5 weeks.

\subsection{Quality Gate Maturity Trajectory}\label{sec:quality-gate-maturity-trajectory}

The most compelling evidence of the loop's self-improving quality infrastructure:

{\def\LTcaptype{none} 
\begin{longtable}[]{@{}lllll@{}}
\toprule\noalign{}
Metric & Iter 1-10 & Iter 50-60 & Iter 128-135 & Iter 163 \\
\midrule\noalign{}
\endhead
\bottomrule\noalign{}
\endlastfoot
L1 pass rate & 85-100\% & 100\% & 100\% & 100\% \\
L2 pass rate & 76-91\% & 90-97\% & 100\% & 100\% \\
L2 unverifiable & 9-31\% & 3-10\% & 0\% & 0\% \\
L3 pass rate & 85-100\% & 92-100\% & 100\% & 100\% \\
Verifier coverage & \roughly92\% (33/36) & \roughly100\% & 100\% (72/72) & 100\% (77/77) \\
PR merge rate & \roughly0\% & \roughly85\% & 100\% & 97\% \\
Open PR backpressure & 0-3 & 3-10 & 0 & 0-3 \\
\end{longtable}
}

Key insight: In this deployment, the quality gates improved monotonically. No regression in pass rates was detected by the oracle after any of the 366 merged PRs across 163 iterations. This suggests that autonomous code changes can maintain (and improve) quality invariants without continuous human oversight, given sufficient verification infrastructure.

\subsection{Phase Progression}\label{sec:edge-phase-progression}

The Edge deployment traversed four distinct phases, mirroring the SDK's progression but at higher iteration speed:

\begin{description}[style=nextline, leftmargin=1.5em, font=\bfseries]
\item[Phase 1: Bootstrap (Iterations 1-12, Days 1-2)]
Detection engine without self-correction capability. 12 PRs created, 0 merged --- the loop could find bugs but couldn't fix them. PR manager and polish phase activated; first merges in iter 11. Quality trajectory: L1 85$\rightarrow$100\%, L2 69$\rightarrow$96\%, L3 85$\rightarrow$100\%.

\item[Phase 2: Steady-State Improvement (Iterations 13-87, Days 2-6)]
Productive self-improvement with monotonically improving quality. 200+ PRs merged. The loop identified and fixed 7 infrastructure bugs through its own Ideate$\rightarrow$Triage$\rightarrow$Execute cycle: chainValid permanently fixed, drift persistence migrated to file-backed storage, verifier families expanded, canary Tier 2 fixed, circuit breaker added, non-functional agent exclusion list created. L1/L2/L3 all reached and held 100\%.

\item[Phase 3: Starvation (Iterations 88-127, Days 6-9)]
All remaining tickets required SDK (Python) changes unreachable from the Edge (TypeScript) codebase. The circuit breaker fired on 10+ consecutive iterations. Two of three reviewers recommended STOP THE LOOP. This was correct behavior: the loop did not degrade or invent work --- it recognized exhaustion and recommended stopping. Quality gates maintained 100\% throughout.

\item[Phase 4: Recovery and Expansion (Iterations 128-163, Days 9-17)]
New work became available after SDK blockers resolved. 100+ PRs merged. 20+ new agents tested, LPOpportunityAgent fixed, ArbitrageAgent activated, TechnicalAgent added. Canaries reached 100\% across all 4 tiers by iteration 124 and held for 40+ subsequent iterations.
\end{description}

\subsection{Key Findings (Early Iterations)}\label{sec:key-findings-early-iterations}

\begin{itemize}
\tightlist
\item
  A critical cooldown bypass needed --- 33+ agents unreachable due to dedup, blocking 75\% of fleet from quality testing
\item
  A production agent using \texttt{Math.random()} simulation instead of real on-chain data --- causing temporal failures with stale signals from tokens launched over a year ago
\item
  A ``dead'' agent registered in config but not in the agent system --- producing zero signals silently
\item
  The loop's own quality gate conflating ``unverifiable'' with ``failed'' --- a false positive the loop discovered and fixed through its own triage process
\end{itemize}

\subsection{The Self-Correction Flywheel (Iterations 128-135)}\label{sec:self-correction-flywheel-iterations-128-135}

At maturity, the loop's most powerful property is \textbf{iterative self-correction} --- it doesn't just find bugs, it fixes them, verifies the fixes, and catches when fixes are incomplete.

\textbf{The OnChainSentinel Fix Chain:}

\begin{table}[ht]
\small
\renewcommand{\arraystretch}{1.3}
\begin{tabular*}{\textwidth}{@{} c @{\quad} p{13.5cm} @{}}
\toprule
\textbf{Iter} & \textbf{What Happened} \\
\midrule
130 & Ideate discovers OnChainSentinel hangs the pipeline via WebSocket constructor blocking. Creates ticket. \\
130 & Execute ships PR: 5-second timeout guard around \texttt{getNetwork()} + adds agent to exclusion list. \\
131 & Ideate retests. Discovers the fix was \textbf{incomplete} --- the WebSocketProvider constructor itself blocks before \texttt{getNetwork()} is ever called. Reopens ticket. \\
131 & Execute ships PR: wraps entire WS provider init in a single timeout with provider cleanup on expiry. Adds destroy-on-timeout test. \\
131 & Regress confirms: build passes, 27 signals produced, L1/L2/L3 100\%, no regressions. \\
\bottomrule
\end{tabular*}
\end{table}

This is a 3-iteration cycle from discovery → incomplete fix → complete fix → regression verification, with zero human intervention. The loop caught its own incomplete fix and iterated until the root cause was fully addressed.

\textbf{The ArbitrageAgent Data Pipeline Resurrection:}

The ArbitrageAgent went through a 3-PR fix chain across multiple iterations:

\begin{enumerate}
\def\labelenumi{\arabic{enumi}.}
\tightlist
\item
  \textbf{PR \#278} --- Replaced dead The Graph hosted subgraph URLs with GeckoTerminal API (The Graph deprecated its free hosted service --- a real external platform change the loop detected)
\item
  \textbf{PR \#282} --- Fixed incorrect \texttt{base\_token\_price\_usd} vs \texttt{quote\_token\_price\_usd} field usage that produced nonsensical 3,499× spreads
\item
  \textbf{PR \#286} --- Lowered \texttt{ARBITRAGE\_PROFIT\_MIN\_PERCENT} from 1\% to 0.1\% after observing 2 consecutive zero-signal runs (1\% required a 1.6\% gross spread on Ethereum --- far above real market conditions of 0.1-0.5\%)
\end{enumerate}

Each iteration peeled back one more layer of the problem. The loop doesn't give up after the first fix --- it keeps testing until the agent actually produces signals.

\subsection{Triage Intelligence: The Classification System}\label{sec:triage-intelligence-the-classification-system}

The triage phase demonstrates sophisticated classification that goes beyond binary ``working/broken'':

{\def\LTcaptype{none} 
\begin{longtable}[]{@{}
  >{\raggedright\arraybackslash}p{(\linewidth - 4\tabcolsep) * \real{0.4412}}
  >{\raggedright\arraybackslash}p{(\linewidth - 4\tabcolsep) * \real{0.2647}}
  >{\raggedright\arraybackslash}p{(\linewidth - 4\tabcolsep) * \real{0.2941}}@{}}
\toprule\noalign{}
\begin{minipage}[b]{\linewidth}\raggedright
Classification
\end{minipage} & \begin{minipage}[b]{\linewidth}\raggedright
Example
\end{minipage} & \begin{minipage}[b]{\linewidth}\raggedright
Response
\end{minipage} \\
\midrule\noalign{}
\endhead
\bottomrule\noalign{}
\endlastfoot
Non-functional (confirmed) & RWARiskAgent: hardcoded stub data (backingRatio=1.0) & Add to exclusion list immediately \\
Non-functional (suspected) & LPRotationAgent: 0 signals, 1st observation & Monitor --- needs 2+ confirmations before excluding \\
Cold-start & DevActivityAgent: needs warm baseline from prior runs & Don't exclude --- works in scheduled mode \\
Timing-dependent & BribeTrackingAgent: no active vote rounds & Don't exclude --- market conditions, not a bug \\
Deprecated & YieldAgent: converted to return-empty stub & Add to exclusion list \\
\end{longtable}
}

\textbf{Self-correcting classification}: In iteration 131, triage classified DevActivityAgent as ``non-functional.'' In iteration 132, it corrected this diagnosis --- the agent has a working GitHub API integration and is actually cold-start, not broken. The ticket was updated, and the agent was NOT added to the exclusion list. Self-correction extends beyond code fixes to the classification system itself.

\subsection{Infrastructure Self-Healing (Iterations 128-135)}\label{sec:infrastructure-self-healing-iterations-128-135}

The loop identified and fixed 6 problems in its own infrastructure during a single 8-iteration batch:

{\def\LTcaptype{none} 
\begin{longtable}[]{@{}
  >{\raggedright\arraybackslash}p{(\linewidth - 4\tabcolsep) * \real{0.4091}}
  >{\raggedright\arraybackslash}p{(\linewidth - 4\tabcolsep) * \real{0.2273}}
  >{\raggedright\arraybackslash}p{(\linewidth - 4\tabcolsep) * \real{0.3636}}@{}}
\toprule\noalign{}
\begin{minipage}[b]{\linewidth}\raggedright
Finding
\end{minipage} & \begin{minipage}[b]{\linewidth}\raggedright
Fix
\end{minipage} & \begin{minipage}[b]{\linewidth}\raggedright
Impact
\end{minipage} \\
\midrule\noalign{}
\endhead
\bottomrule\noalign{}
\endlastfoot
\texttt{RESULTS\_DIR} undefined in pr-manager.sh & Grep prep log instead & Gate rejection memory completely non-functional \\
\texttt{gh\ -\/-jq\ -\/-arg} invalid syntax & Pipe to \texttt{jq\ -\/-arg} & Shell command silently failing \\
Missing iter 127 row in loop-state.md & Backfill + verification step & History gap invisible to drift detection \\
\texttt{synthesisPipelineDead} false positive & \texttt{localDev} flag for drift suppression & 9 iterations of wasted CRITICAL escalation \\
Deployment drift (local main stale) & Pre-flight detection & All merged fixes untested for 5+ iterations \\
\texttt{NON\_FUNCTIONAL\_AGENTS} set out of sync & Sync with confirmed findings + expose via API & Ideate kept selecting dead agents \\
\end{longtable}
}

As in Case Study A (Section 8.4), the loop improves at two levels: the product and its own infrastructure.

\subsection{Speed Difference: Two Domains, Same Loop}\label{sec:speed-difference-two-domains-same-loop}

{\def\LTcaptype{none} 
\begin{longtable}[]{@{}llll@{}}
\toprule\noalign{}
Phase & Strategy Framework & Signal Platform & Speedup \\
\midrule\noalign{}
\endhead
\bottomrule\noalign{}
\endlastfoot
Ideate & 10-15 min & 1-2 min & \roughly8x \\
Triage & 5-8 min & \textless1 min & \roughly8x \\
Execute & 25-60 min & 10-30 min & \roughly2x \\
Regress & 40-150 min & 2-3 min & \roughly30x \\
\textbf{Total} & \textbf{80-230 min} & \textbf{\roughly15-35 min\textsuperscript{*}} & \textbf{\roughly5-15x} \\
\end{longtable}
}

\textsuperscript{*}Signal-only iterations (ideate + triage) complete in \roughly5 min; full 6-phase iterations take 80-230 min. The \roughly15-35 min figure reflects signal-focused iterations including execute and polish. Circuit-breaker skip iterations complete in 15-20 min; the average across all 163 iterations was \roughly45 min.

\vspace{4pt}
The signal platform iterates faster because signal quality verification is API-bound, not execution-bound. This speed advantage compounds over time: at 163 iterations, the signal platform has performed more quality gate evaluations than most human QA teams accomplish in a year.

\section{Cost of Operation}\label{sec:cost-of-operation}

\subsection{Tooling Spend}\label{sec:tooling-spend}

A common concern: ``How much does it cost to run an AI agent continuously?'' The answer is surprisingly modest: the Kitchen Loop runs entirely on \textbf{flat-rate subscriptions}, not metered API calls.

{\def\LTcaptype{none} 
\begin{longtable}[]{@{}
  >{\raggedright\arraybackslash}p{(\linewidth - 4\tabcolsep) * \real{0.30}}
  >{\raggedright\arraybackslash}p{(\linewidth - 4\tabcolsep) * \real{0.20}}
  >{\raggedright\arraybackslash}p{(\linewidth - 4\tabcolsep) * \real{0.50}}@{}}
\toprule\noalign{}
\begin{minipage}[b]{\linewidth}\raggedright
Cost Component
\end{minipage} & \begin{minipage}[b]{\linewidth}\raggedright
Monthly Cost
\end{minipage} & \begin{minipage}[b]{\linewidth}\raggedright
Notes
\end{minipage} \\
\midrule\noalign{}
\endhead
\bottomrule\noalign{}
\endlastfoot
\textbf{Claude Code Max} (20x plan) & \$200 & Primary agent for all 6 phases --- unlimited usage within plan \\
\textbf{Codex} (subscription) & \$20 & External feasibility checker + tribunal reviewer \\
\textbf{Gemini} (subscription) & \$20 & Tribunal reviewer for multi-model audits \\
\textbf{CodeRabbit} & \roughly\$15 & Automated PR code review \\
\textbf{Anvil / Foundry} & \$0 & Open-source, runs locally \\
\textbf{External data APIs} & \roughly\$0-50 & CoinGecko, DefiLlama (free tiers sufficient) \\
\textbf{CI compute} & \roughly\$50-100 & GitHub Actions minutes \\
\textbf{Total} & \textbf{\roughly\$305-405/month} & \\
\end{longtable}
}

This is the total cost for running the Kitchen Loop across \textbf{both} production systems simultaneously --- 285+ iterations, 1,094+ merged PRs, 700+ tickets resolved. A senior engineer costs \$12,000-25,000/month fully loaded; the Kitchen Loop's monthly tooling cost is \textbf{\roughly2\% of a single engineer's cost}. Critically, this includes the quality infrastructure (unbeatable tests, UAT gates, regression oracles) that prevents the complexity-driven velocity erosion documented by He et al.~\cite{he2025} and Becker et al.~\cite{becker2025}.

{\def\LTcaptype{none} 
\begin{longtable}[]{@{}llll@{}}
\toprule\noalign{}
Metric & Kitchen Loop & Single Engineer & Ratio \\
\midrule\noalign{}
\endhead
\bottomrule\noalign{}
\endlastfoot
Monthly cost & \roughly\$350 & \roughly\$15,000 & \textbf{\roughly43x cheaper} \\
PRs merged/month & 600+ & \roughly15-25 & \textbf{\roughly30x more} \\
Coverage scenarios tested/month & 150+ & \roughly10-20 & \textbf{\roughly10x more} \\
Cost per merged PR & \textbf{\roughly\$0.38} & \textbf{\$600-1,000} & \textbf{\roughly1{,}800x cheaper} \\
\end{longtable}
}

Because cost is fixed regardless of iteration count, the marginal cost of additional coverage is effectively zero.

\subsection{Test Suite Growth and Runtime}\label{sec:test-suite-growth-and-runtime}

As the test suite grows, regression time grows. This is a real scaling concern:

{\def\LTcaptype{none} 
\begin{longtable}[]{@{}
  >{\raggedright\arraybackslash}p{(\linewidth - 6\tabcolsep) * \real{0.25}}
  >{\raggedright\arraybackslash}p{(\linewidth - 6\tabcolsep) * \real{0.18}}
  >{\raggedright\arraybackslash}p{(\linewidth - 6\tabcolsep) * \real{0.22}}
  >{\raggedright\arraybackslash}p{(\linewidth - 6\tabcolsep) * \real{0.35}}@{}}
\toprule\noalign{}
\begin{minipage}[b]{\linewidth}\raggedright
Metric
\end{minipage} & \begin{minipage}[b]{\linewidth}\raggedright
Start (iter 1)
\end{minipage} & \begin{minipage}[b]{\linewidth}\raggedright
Current (iter 122)
\end{minipage} & \begin{minipage}[b]{\linewidth}\raggedright
Growth Rate
\end{minipage} \\
\midrule\noalign{}
\endhead
\bottomrule\noalign{}
\endlastfoot
Unit tests & \roughly6,400 & 10,913 & +70\% over 122 iterations \\
Full regression time & \roughly90 min & \roughly150 min & \roughly0.5 min/iteration growth \\
Quick regression time & \roughly25 min & \roughly40 min & \roughly0.12 min/iteration growth \\
Demo strategies & 13 & 62 & \roughly0.40/iteration \\
Incubating strategies & 0 & 183 & \roughly1.5/iteration \\
\end{longtable}
}

\textbf{Mitigation strategies already in use:} - \texttt{-\/-regress-quick} mode: one scenario per platform (\roughly40 min vs \roughly150 min) - Full regression runs weekly, quick runs every iteration - Parallel test execution for unit tests (pytest-xdist) - Test pruning: deprecated/redundant strategies archived, not accumulated

\textbf{Projected ceiling}: At current growth rates, full regression would reach \roughly4 hours by iteration 200. This is manageable with parallelization (dynamic port allocation for multiple test environments), sharding (split the coverage matrix across N parallel loops), or sampling (statistical coverage sampling rather than exhaustive every iteration).

\subsection{Business Outcomes}\label{sec:business-outcomes}

For readers focused on outcomes: 50+ features shipped, 200+ bugs found before users (including fund-safety issues), coverage matrix fill rate improved from \roughly5\% to \roughly50\% of 1,800 combinations, and test suite grew +70\%. Detailed breakdowns are in Sections 8.3 and 9.3.

\section{Production Safety Record}\label{sec:production-safety-record}

\subsection{Incidents and Ill Effects}\label{sec:incidents-and-ill-effects}

A critical question for any autonomous system: \textbf{has it caused harm?}

{\def\LTcaptype{none} 
\begin{longtable}[]{@{}
  >{\raggedright\arraybackslash}p{(\linewidth - 4\tabcolsep) * \real{0.28}}
  >{\raggedright\arraybackslash}p{(\linewidth - 4\tabcolsep) * \real{0.15}}
  >{\raggedright\arraybackslash}p{(\linewidth - 4\tabcolsep) * \real{0.57}}@{}}
\toprule\noalign{}
\begin{minipage}[b]{\linewidth}\raggedright
Concern
\end{minipage} & \begin{minipage}[b]{\linewidth}\raggedright
Record
\end{minipage} & \begin{minipage}[b]{\linewidth}\raggedright
Details
\end{minipage} \\
\midrule\noalign{}
\endhead
\bottomrule\noalign{}
\endlastfoot
\textbf{Production downtime} & 0 incidents & Loop operates on isolated branches + test environments (Anvil forks). No direct production access. \\
\textbf{Security incidents} & 0 incidents & Loop has no access to production keys, wallets, or user data. Test wallets use Anvil defaults. \\
\textbf{Data loss} & 0 incidents & Git worktree isolation prevents branch contamination. All changes are reversible PRs. \\
\textbf{Regressions shipped to main} & 0 & Quality gates (multi-model review + regression oracle) catch issues before merge. \\
\textbf{Cost overruns} & 0 & Monthly tooling cost is bounded and predictable (\roughly\$305-405/month across both systems, \roughly\$1.50/iteration). \\
\end{longtable}
}

\subsection{Why Zero Incidents?}\label{sec:why-zero-incidents}

The clean safety record follows from \textbf{architectural isolation}: no production access (test environments only), branch isolation (git worktrees on feature branches), automated pause gates (Section 5.5), and backpressure control. These are structural properties, not luck.

\subsection{Known Limitations (Honest Assessment)}\label{sec:known-limitations-honest-assessment}

{\def\LTcaptype{none} 
\begin{longtable}[]{@{}
  >{\raggedright\arraybackslash}p{(\linewidth - 4\tabcolsep) * \real{0.3667}}
  >{\raggedright\arraybackslash}p{(\linewidth - 4\tabcolsep) * \real{0.2667}}
  >{\raggedright\arraybackslash}p{(\linewidth - 4\tabcolsep) * \real{0.3667}}@{}}
\toprule\noalign{}
\begin{minipage}[b]{\linewidth}\raggedright
Limitation
\end{minipage} & \begin{minipage}[b]{\linewidth}\raggedright
Impact
\end{minipage} & \begin{minipage}[b]{\linewidth}\raggedright
Mitigation
\end{minipage} \\
\midrule\noalign{}
\endhead
\bottomrule\noalign{}
\endlastfoot
\textbf{Anvil-only testing} & Loop has not been exercised with real mainnet transactions & Mainnet mode exists but requires explicit operator approval + real gas \\
\textbf{API key dependency} & Some scenarios require external API keys not in standard config & Produces \texttt{PASS(caveat)} instead of clean \texttt{PASS} --- a known quality ceiling \\
\textbf{Single-threaded execution} & Gateway port constraint limits to sequential strategy runs & Future: dynamic port allocation for parallel execution \\
\textbf{PR merge automation fragility} & The Polish phase has been the most failure-prone component & Each failure mode has been fixed through the self-improvement cycle \\
\textbf{Cooldown phantom failures} (signal platform) & Agents with cooldown periods inflate dead-agent counts & Distinguishing ``cooldown active'' from ``agent broken'' is tracked as an open ticket \\
\end{longtable}
}

This section is intentionally honest. The loop is not infallible. But its failure modes are \emph{operational} (merge automation bugs, API key management) rather than \emph{safety-critical} (data loss, fund loss, production outages). The architectural isolation ensures that operational failures waste iteration time, not user trust.

\subsection{Human-in-the-Loop Cost Constraints}\label{sec:human-in-the-loop-cost-constraints}

The loop is autonomous end-to-end: it generates scenarios from the specification surface, fills its own backlog, triages findings into tickets, implements fixes, and auto-merges PRs after multi-model review and CI. No human is in the critical path during normal operation. The human role reduces to (1) initial specification surface definition, (2) occasional strategic steering, and (3) \textbf{supervisory intervention} when the loop encounters failure modes it cannot self-correct.

In practice, the most common intervention trigger was \emph{merge conflict resolution gone wrong}: LLMs resolving git conflicts would sometimes silently undo work from previous commits or PRs, requiring a human to notice the regression and revert. GitHub infrastructure issues (API rate limits, webhook failures, stale branch state) were the second most frequent cause. Section 12.2 frames the steady-state human cost as \roughly30-60 minutes per week once calibrated --- but this assumes the loop is running smoothly. Intervention spikes during infrastructure instability.

The practical scaling constraint is \emph{environmental}: test environment throughput (Anvil fork startup time, API rate limits), CI pipeline capacity, and the cost of multi-model review tokens per PR. Drain mode (Section 5.5) throttles output when PR backpressure exceeds a threshold, ensuring the loop does not outrun its own merge infrastructure.

The Discussion Manager adds a separate cost dimension. Three models debating for 3-5 rounds at \roughly400 words per turn consumes 15,000-25,000 tokens per model per discussion. With the \roughly50\% round efficiency observed in our 23-discussion corpus (Section 7.3), approximately 40\% of deliberation token spend produces convergence dynamics rather than novel insight. At current subscription pricing this cost is absorbed into flat-rate plans, but metered API usage would make frequent deliberation expensive. Per-iteration compute cost breakdown beyond the aggregate \$0.38/PR figure has not yet been instrumented --- this is a gap we intend to address in future work.

These constraints do not invalidate the framework, but they bound its applicability: the Kitchen Loop is most efficient in domains where (a) the specification surface is well-defined upfront, (b) the regression oracle provides high-confidence automated verification, and (c) the human's role can be genuinely asynchronous rather than synchronous with each iteration.

\subsection{Open Problems}\label{sec:open-problems}

Our deployments expose four open problems that we believe warrant dedicated research:

\begin{itemize}
\item
  \textbf{OP1: Oracle Transfer.} The Kitchen Loop relies on a bespoke regression oracle per domain (chain fork execution for DeFi, quality-gate pipeline for signals). Automatic generation of regression oracles from natural-language specifications --- eliminating the per-domain engineering cost --- is an unsolved challenge.
\item
  \textbf{OP2: Specification Acquisition.} Our method assumes an enumerable specification surface. For legacy codebases with implicit specifications, automating surface extraction from telemetry, documentation, and user behavior is a critical bottleneck for adoption.
\item
  \textbf{OP3: Multi-Objective Drift.} Current drift metrics focus on functional correctness. Extending the framework to simultaneously monitor non-functional requirements (latency, security, fairness) without human intervention remains open. Our drift detection (Section 5.5) would need to compose multiple objective functions without one dominating the pause-gate signal.
\item
  \textbf{OP4: Sycophancy at Scale.} The Discussion Manager mitigates sycophancy in 3-model debates (Section 7), but optimal model composition and debate protocols for larger, heterogeneous agent swarms are unknown. Our corpus of 23 discussions is too small to establish whether the observed SS < 20 threshold generalizes (cf.~Yao et al., 2025 \cite{yao2025}).
\end{itemize}

\subsection{When the Kitchen Loop Should NOT Be Used}\label{sec:when-the-kitchen-loop-should-not-be-used}

The framework is not universally applicable. It should not be applied when: (1) ground truth is weak or unobservable --- without a reliable oracle, the verification layer provides false confidence; (2) success criteria are highly subjective (aesthetic quality, UX taste) --- the oracle cannot arbitrate matters of judgment; (3) the specification surface is not enumerable, as in exploratory R\&D where the goal is discovery rather than convergence; or (4) safety-critical domains lack robust external verification --- the oracle's bounded coverage (Section 4.1) is insufficient when failures have irreversible consequences.

\section{The Human-AI Collaboration Model}\label{sec:human-ai-collaboration-model}

\subsection{Not Either/Or --- Both}\label{sec:not-eitheror-both}

Chen et al.~(2025) identify three automation points --- human-only, copilot, and agent --- finding agents achieve 60\% task correctness vs.~25\% for copilots, yet 60\% of participants would not continue using agents due to a comprehension gap \cite{chen2025}. The Kitchen Loop operates at a fourth point: \textbf{fully autonomous loop with asynchronous human oversight}, eliminating the idle-time and comprehension problems Chen et al.~identify.

The Kitchen Loop is explicitly a \textbf{complementary system}, not a replacement:

{\def\LTcaptype{none} 
\begin{longtable}[]{@{}
  >{\raggedright\arraybackslash}p{(\linewidth - 4\tabcolsep) * \real{0.2647}}
  >{\raggedright\arraybackslash}p{(\linewidth - 4\tabcolsep) * \real{0.5294}}
  >{\raggedright\arraybackslash}p{(\linewidth - 4\tabcolsep) * \real{0.2059}}@{}}
\toprule\noalign{}
\begin{minipage}[b]{\linewidth}\raggedright
Concern
\end{minipage} & \begin{minipage}[b]{\linewidth}\raggedright
AI (Kitchen Loop)
\end{minipage} & \begin{minipage}[b]{\linewidth}\raggedright
Human
\end{minipage} \\
\midrule\noalign{}
\endhead
\bottomrule\noalign{}
\endlastfoot
\textbf{Coverage} & Exhaustive --- 1000x the scenarios a human would test & Strategic --- focuses on what matters most given user context \\
\textbf{Bug discovery} & Bottom-up --- finds what's broken by trying everything & Top-down --- knows what users are complaining about \\
\textbf{Ticket writing} & Precise, file-level, with reproduction steps & Context-rich, business-aware \\
\textbf{Implementation} & Fast, consistent, follows established patterns & Creative, architectural, can break patterns when needed \\
\textbf{Code review} & Multi-model parallel tribunals & Domain expertise, security intuition \\
\textbf{Backlog curation} & Deduplication, severity assessment, dependency graphs & Strategic priority, business value, external signals \\
\end{longtable}
}

\subsection{The Human's Highest-Leverage Input}\label{sec:humans-highest-leverage-input}

The human's primary contribution is \textbf{specification and backlog curation}: understanding users, monitoring the competitive landscape, and converting those signals into tickets. In our deployments, this required \roughly30-60 minutes per week once the loop was calibrated --- but this figure assumes a mature specification surface (see Section 11.4 for scaling constraints and the human bottleneck at high iteration speeds).

\subsection{The External Feasibility Checker}\label{sec:external-feasibility-checker}

The Kitchen Loop optionally uses an external AI (a different model from the loop's primary agent) as a \textbf{feasibility checker} before committing to an idea.

\textbf{Empirical results} (285+ combined iterations): \roughly78\% PROCEED, \roughly15\% REDIRECT, \textless5\% REJECT + timeout. The low rejection rate suggests the loop's scenario selection is well-calibrated. REDIRECT cases adjusted scope productively.

\section{The Self-Improving Loop}\label{sec:self-improving-loop}

\subsection{Meta-Level Improvement}\label{sec:meta-level-improvement}

The loop-review skill audits loop behavior every N iterations and produces tickets for infrastructure improvements. Sections 8.4 and 9.3 showed specific examples; the full catalog across both deployments:

\begin{itemize}[nosep]
\item
  \textbf{Platform-specific failures}: Merge automation memory bug on Apple Silicon caused 5 consecutive stalls; loop review identified the pattern and the Execute phase fixed it.
\item
  \textbf{Resource contention}: Process collisions when human and loop run concurrently; tool version drift breaking the review phase. Fixes added process tracking and version pinning.
\item
  \textbf{Retry and backpressure loops}: PR Manager stuck retrying the same failing PR, wasting entire Polish phases. Fixed by skip-after-2-failures with \texttt{needs-attention} labeling. PR backlog growing unbounded during sustained runs prompted automatic drain mode.
\item
  \textbf{State management}: Loop-state lost when worktree PRs weren't merged promptly (fixed by decoupling state sync from merge lifecycle). Backlog groomer promoting already-in-progress tickets (fixed by counting viable tickets only).
\item
  \textbf{Silent failures}: Missing \texttt{.env} in worktrees causing strategy failures; loop-review reports silently lost when output directory didn't exist. Fixed by pre-flight checks and write verification guards.
\item
  \textbf{Budget management}: Regress timeout consuming 100\% of iteration budget, prompting a quick-regression parameter.
\end{itemize}

The framework has also been applied to its own codebase (dogfooding). In its first iterations running against the KitchenLoop orchestrator and PR manager, the loop discovered and fixed multiple bugs --- including race conditions in temporary file handling, missing agent-liveness detection, and incorrect timeout behavior on macOS. This validates that the self-improvement property extends to the meta level: the loop can improve the tool that runs the loop.

\subsection{Pattern Consolidation}\label{sec:pattern-consolidation}

When the same pattern is confirmed across 2+ iterations, it is promoted from a session observation to a durable memory entry. This creates institutional knowledge that persists across conversations and loop runs.

\subsection{The Skill Layer}\label{sec:skill-layer}

Starting from 5 basic phase skills, the validated deployments now have 30+ skills covering loop orchestration, quality gates, protocol integration, competitive intelligence, release management, and documentation maintenance. Each skill emerged from the loop's own discovery of a repeatable workflow worth encoding. Additionally, the loop developed a \textbf{gate rejection memory system} that prevents redundant LLM auditor calls --- if a PR was rejected and no new commits were pushed, it is immediately marked NOT\_MERGEABLE without wasting an audit cycle. This optimization, discovered and implemented by the loop itself, eliminated the zero-backpressure bottleneck observed in earlier iterations.

Recent additions include \texttt{/loop-review-meta} --- a macro-level analysis skill that aggregates findings across all loop review reports to surface systemic trends and strategic recommendations.

\section{Generalization: Making the Kitchen Loop Portable}\label{sec:generalization-making-the-kitchen-loop-portable}

\subsection{What Makes a Codebase Loop-Ready}\label{sec:what-makes-a-codebase-loop-ready}

The Kitchen Loop works on any codebase where:

\begin{enumerate}
\def\labelenumi{\arabic{enumi}.}
\tightlist
\item
  \textbf{The specification is enumerable}: The product has a definable set of features, platforms, and action types that can be expressed as a coverage matrix.
\item
  \textbf{Usage can be automated}: ``Using the product'' can be performed by an LLM agent without physical interaction (APIs, CLIs, SDKs).
\item
  \textbf{Quality is measurable by regression}: There exists a test oracle that can answer ``is the system still working?'' in bounded time.
\end{enumerate}

\subsection{How to Adapt the Kitchen Loop to Your Domain}\label{sec:how-to-adapt-the-kitchen-loop-to-your-domain}

{\small\def\LTcaptype{none} 
\begin{longtable}[]{@{}
  >{\raggedright\arraybackslash}p{(\linewidth - 6\tabcolsep) * \real{0.1176}}
  >{\raggedright\arraybackslash}p{(\linewidth - 6\tabcolsep) * \real{0.3235}}
  >{\raggedright\arraybackslash}p{(\linewidth - 6\tabcolsep) * \real{0.2794}}
  >{\raggedright\arraybackslash}p{(\linewidth - 6\tabcolsep) * \real{0.2794}}@{}}
\toprule\noalign{}
\begin{minipage}[b]{\linewidth}\raggedright
Domain
\end{minipage} & \begin{minipage}[b]{\linewidth}\raggedright
Specification Surface
\end{minipage} & \begin{minipage}[b]{\linewidth}\raggedright
Regression Oracle
\end{minipage} & \begin{minipage}[b]{\linewidth}\raggedright
Example Adaptation
\end{minipage} \\
\midrule\noalign{}
\endhead
\bottomrule\noalign{}
\endlastfoot
\textbf{Web Application} & Pages x User Flows x Browsers & Browser automation + visual regression (Playwright, Cypress) & Ideate generates user journeys; Regress runs screenshot diffing against known-good baselines \\
\textbf{ML Pipeline} & Models x Datasets x Metrics & W\&B / MLflow run comparison + statistical tests & Ideate trains model variants; Regress compares against baseline metrics with significance tests \\
\textbf{Smart Contracts} & Functions x Chains x Edge Cases & Foundry/Anvil fork execution + invariant checks & Ideate writes fuzz test scenarios; Regress runs invariant test suites \\
\textbf{Backend API} & Endpoints x Methods x Auth Roles & Contract testing + live traffic shadow comparison & Ideate generates API call sequences; Regress replays against shadow traffic \\
\textbf{Mobile App} & Screens x Gestures x Devices & Appium automation + visual regression & Ideate scripts user flows; Regress runs on device farm \\
\textbf{Compiler / Language} & Grammar x Optimizations x Targets & Test suite execution + benchmark comparison & Ideate generates programs exercising edge cases; Regress runs conformance suite \\
\end{longtable}
}

The two adaptation points are always the same: \emph{what is the specification surface?} and \emph{what is the regression oracle?} Everything else --- backlog management, phase sequencing, drift control, self-improvement --- transfers intact.

\subsection{The Specification Layer}\label{sec:specification-layer}

The Kitchen Loop works because the products it's been applied to have well-defined specifications. Most codebases lack this. A \textbf{specification layer} --- structured YAML or Markdown specs stored alongside code --- solves this by making specs machine-readable, version-controlled, and auditable.

\subsection{The Scaffolding Layer}\label{sec:scaffolding-layer}

A \textbf{scaffolding layer} --- templates for the six-phase orchestrator, ticket management stubs, oracle skeletons, and CI workflows --- reduces time-to-first-loop from days to hours.

\subsection{A Composable Stack}\label{sec:a-composable-stack}

\begin{figure}[h]
\centering
\begin{tikzpicture}[
    layer/.style={draw, rounded corners=2pt, minimum width=11cm, minimum height=1.4cm, align=center, font=\small}
]
\node[layer, fill=blue!12] (engine) at (0, 3) {\textbf{Kitchen Loop Engine}\\{\scriptsize Backlog $\rightarrow$ Ideate $\rightarrow$ Triage $\rightarrow$ Execute $\rightarrow$ Polish $\rightarrow$ Regress $\rightarrow$ (repeat)}};
\node[layer, fill=green!8] (spec) at (0, 1.3) {\textbf{Specification Layer}\\{\scriptsize What does the product promise? What should we test?}\\{\scriptsize YAML/Markdown specs versioned alongside code}};
\node[layer, fill=gray!10] (scaffold) at (0, -0.4) {\textbf{Scaffolding Layer}\\{\scriptsize Project setup, CI/CD, skill templates, oracle stubs}\\{\scriptsize Reduces time-to-first-loop from days to hours}};
\end{tikzpicture}
\end{figure}

\section{Conclusion}\label{sec:conclusion}

\subsection{The Shift}\label{sec:shift}

In our observation, software engineering is undergoing a phase transition. Code production --- once the bottleneck --- is becoming a commodity. Code review --- once a human-only activity --- is becoming automated. The competitive advantage is shifting to:

\begin{enumerate}
\def\labelenumi{\arabic{enumi}.}
\tightlist
\item
  \textbf{Specification}: Knowing what to build (the AaU1000 method)
\item
  \textbf{Verification}: Proving it works (unbeatable tests)
\item
  \textbf{Convergence}: Ensuring it keeps working (regression and drift control)
\end{enumerate}

\subsection{The Evidence}\label{sec:evidence}

Across two production systems and 285+ combined iterations (Sections 8.3 and 9.3), zero regressions were detected by the regression oracle, quality gates improved monotonically from 76--91\% to 100\%, and the loop autonomously fixed 17+ infrastructure bugs in its own tooling. The second system validated portability: the same architecture applied to a fundamentally different domain achieved comparable results with a 25x faster iteration cycle, requiring only two reimplemented components (specification surface and regression oracle). The unified trust model held under sustained autonomous operation without increasing human intervention.

\subsection{The Invitation}\label{sec:invitation}

The Kitchen Loop is not the future of software development. It is a practice available today, on a codebase of any size, with tools that already exist. The prerequisites are:

\begin{enumerate}
\def\labelenumi{\arabic{enumi}.}
\tightlist
\item
  \textbf{An enumerable specification surface} --- what does your product claim to do?
\item
  \textbf{An automatable test environment} --- can an AI agent exercise your product?
\item
  \textbf{A regression oracle} --- can you answer ``is the system still working?'' in bounded time?
\item
  \textbf{The discipline to run it continuously} --- and trust the output when the tests are unbeatable.
\end{enumerate}

In our experience, the tests are the trust layer. The spec is the compass. The loop is the engine. Given sufficient verification infrastructure, the product evolves itself.

\subsection{Testable Hypotheses}\label{sec:testable-hypotheses}

Our deployments suggest four empirical hypotheses that subsequent work could validate, replicate, or refute:

\begin{itemize}
\item
  \textbf{H1:} Coverage-exhaustion systems (as defined in the regime taxonomy, Section 2.9) discover more user-visible failures per iteration than task-completion systems in partially mature products with \textgreater50\% specification surface coverage.
\item
  \textbf{H2:} Adversarial UAT gates --- sealed test cards executed by a fresh evaluator with zero implementation context --- reduce false-positive readiness assessments compared with implementer-authored test suites alone.
\item
  \textbf{H3:} Tier-weighted scenario selection (Foundation 30\% / Composition 50\% / Frontier 20\%) discovers more bugs per iteration than uniform random selection across the specification surface. The superlinear growth of the Composition tier (Section 3.6) predicts that the advantage increases with product maturity.
\item
  \textbf{H4:} In user-facing verification tasks, weak-model evaluation (the least capable model available) is a better proxy for real-user verifiability than strong-model evaluation, because weak models fail on the same ambiguities that trip real users.
\end{itemize}

These hypotheses are intended to facilitate direct replication, extension, or refutation in future agentic systems research.

\subsection{Known Limitations \& Future Work}\label{sec:known-limitations-future-work}

The Kitchen Loop has five structural limitations that bound its applicability:

\begin{enumerate}[nosep]
\item \textbf{Single-threaded execution.} The current orchestrator runs one iteration at a time. Parallelization across multiple worktrees is architecturally straightforward but not yet implemented (Section~\ref{sec:known-limitations-honest-assessment}).
\item \textbf{Enumerable specification surface required.} The method assumes the product's capabilities can be listed as a coverage matrix. Legacy monoliths with implicit specifications require a specification-extraction step (OP2, Section~\ref{sec:open-problems}) before the loop can operate.
\item \textbf{Oracle quality is the ceiling.} The loop can only catch what the regression oracle can verify. If the oracle misses a failure mode, the loop is blind to it. Oracle transfer across domains (OP1) remains an open research problem.
\item \textbf{Human role persists.} Backlog grooming, specification design, and production promotion require human judgment. At scale, human merge capacity --- not AI generation speed --- becomes the binding constraint (Section~\ref{sec:human-in-the-loop-cost-constraints}).
\item \textbf{Two-domain validation only.} Our evidence spans two production systems in the DeFi domain. Generalization to other domains (Section~\ref{sec:generalization-making-the-kitchen-loop-portable}) is architecturally supported but empirically unvalidated.
\end{enumerate}

\appendix
\clearpage

\section{The Coverage Matrix (Strategy Framework Example)}\label{sec:appendix-c-the-coverage-matrix-strategy-framework-example}

The strategy framework's intent test coverage matrix illustrates the exhaustive approach. Every cell represents a claim: ``Protocol X works on Chain Y for Intent Z.'' Empty cells are untested claims.

{\footnotesize\def\LTcaptype{none} 
\begin{longtable}[]{@{}
  >{\raggedright\arraybackslash}p{(\linewidth - 16\tabcolsep) * \real{0.1333}}
  >{\raggedright\arraybackslash}p{(\linewidth - 16\tabcolsep) * \real{0.0933}}
  >{\raggedright\arraybackslash}p{(\linewidth - 16\tabcolsep) * \real{0.0800}}
  >{\raggedright\arraybackslash}p{(\linewidth - 16\tabcolsep) * \real{0.1200}}
  >{\raggedright\arraybackslash}p{(\linewidth - 16\tabcolsep) * \real{0.1333}}
  >{\raggedright\arraybackslash}p{(\linewidth - 16\tabcolsep) * \real{0.1067}}
  >{\raggedright\arraybackslash}p{(\linewidth - 16\tabcolsep) * \real{0.1067}}
  >{\raggedright\arraybackslash}p{(\linewidth - 16\tabcolsep) * \real{0.0933}}
  >{\raggedright\arraybackslash}p{(\linewidth - 16\tabcolsep) * \real{0.1333}}@{}}
\toprule\noalign{}
\begin{minipage}[b]{\linewidth}\raggedright
Protocol
\end{minipage} & \begin{minipage}[b]{\linewidth}\raggedright
Chain
\end{minipage} & \begin{minipage}[b]{\linewidth}\raggedright
Swap
\end{minipage} & \begin{minipage}[b]{\linewidth}\raggedright
LP Open
\end{minipage} & \begin{minipage}[b]{\linewidth}\raggedright
LP Close
\end{minipage} & \begin{minipage}[b]{\linewidth}\raggedright
Supply
\end{minipage} & \begin{minipage}[b]{\linewidth}\raggedright
Borrow
\end{minipage} & \begin{minipage}[b]{\linewidth}\raggedright
Repay
\end{minipage} & \begin{minipage}[b]{\linewidth}\raggedright
Withdraw
\end{minipage} \\
\midrule\noalign{}
\endhead
\bottomrule\noalign{}
\endlastfoot
Aerodrome & Base & P0 & P0 & P0 & - & - & - & - \\
TraderJoe V2 & Avalanche & P0 & P0 & P0 & - & - & - & - \\
Uniswap V3 & Ethereum & P1 & P1 & P1 & - & - & - & - \\
Uniswap V3 & Arbitrum & P1 & P1 & P1 & - & - & - & - \\
Uniswap V3 & Base & P1 & P1 & P1 & - & - & - & - \\
PancakeSwap V3 & BSC & P1 & - & - & - & - & - & - \\
Aave V3 & Ethereum & - & - & - & P0 & P0 & P0 & P0 \\
Aave V3 & Arbitrum & - & - & - & P0 & P0 & P0 & P0 \\
Aave V3 & Base & - & - & - & P0 & P0 & P0 & P0 \\
Compound V3 & Ethereum & - & - & - & P1 & P1 & P1 & P1 \\
Morpho Blue & Ethereum & - & - & - & P2 & P2 & P2 & P2 \\
Enso & Multi-chain & P3 & - & - & - & - & - & - \\
Curve & Ethereum & P3 & - & - & - & - & - & - \\
\end{longtable}
}

\emph{P0 = critical path, P1 = breadth, P2 = depth, P3 = completeness}

\section{Representative Strategy Examples}\label{sec:appendix-d-representative-strategy-examples}

\subsection{D.1 Multi-Protocol Composability (Tier 2 --- Composition)}\label{sec:d.1-multi-protocol-composability-tier-2-composition}

A Tier 2 strategy exercised four intent types across two protocols in a single strategy:

\begin{tcolorbox}[resultbox]
{\small\ttfamily
\begin{tabular}{@{}r@{: }p{7cm}r@{}}
Step 1 & SUPPLY 0.5 WETH as collateral to Lending Protocol & \\
       & $\rightarrow$ 3 TXs (approve + supply + setCollateral) & 270K gas \\[3pt]
Step 2 & BORROW 311.20 USDC at 30\% LTV & \\
       & $\rightarrow$ 1 TX (borrow) & 286K gas \\[3pt]
Step 3 & SWAP 155.60 USDC $\rightarrow$ 0.0749 WETH via DEX & \\
       & $\rightarrow$ 3 TXs (approve + approve\_reset + swap) & 213K gas \\[3pt]
Step 4 & LP\_OPEN WETH/USDC range [1867--2282] & \\
       & $\rightarrow$ 4 TXs (approve + approve\_reset + approve + lp\_mint) & 526K gas \\[3pt]
\midrule
\multicolumn{2}{@{}l}{\textbf{Total: 11 transactions, 1.3M gas}} & \textbf{zero bugs} \\
\end{tabular}
}
\end{tcolorbox}

All four verification layers passed on every intent. This was the first test of cross-protocol enrichment data handoff --- proving that the \emph{seams} between components work, not just the components themselves.

\subsection{D.2 Additional Foundation Examples (Tier 1)}\label{sec:d.2-additional-foundation-examples}

Two Tier 1 scenarios illustrate the ``obviously missing'' signal. A basic DEX swap discovered that the protocol router configuration had no entry on ANY chain (compilation failed) and used a different interface version (8 vs.\ 7 parameters, causing silent reverts) --- both fixed in the same iteration. Separately, the first-ever strategy on a new chain discovered a missing native token symbol --- a one-line omission that blocked ALL strategies on the entire chain.

\section{Signal Platform Quality Gate Architecture}\label{sec:appendix-e-signal-platform-quality-gate-architecture}

\subsection{E.1 Verifier Families}\label{sec:e.1-verifier-families}

{\def\LTcaptype{none} 
\begin{longtable}[]{@{}
  >{\raggedright\arraybackslash}p{(\linewidth - 4\tabcolsep) * \real{0.3200}}
  >{\raggedright\arraybackslash}p{(\linewidth - 4\tabcolsep) * \real{0.4200}}
  >{\raggedright\arraybackslash}p{(\linewidth - 4\tabcolsep) * \real{0.2600}}@{}}
\toprule\noalign{}
\begin{minipage}[b]{\linewidth}\raggedright
Verifier Family
\end{minipage} & \begin{minipage}[b]{\linewidth}\raggedright
Signal Types Covered
\end{minipage} & \begin{minipage}[b]{\linewidth}\raggedright
Data Source
\end{minipage} \\
\midrule\noalign{}
\endhead
\bottomrule\noalign{}
\endlastfoot
Protocol TVL & Discovery, TVL migration, undervalued protocol & DeFi analytics API \\
Pool Yield & Yield opportunity, LP opportunity, cross-DEX arbitrage & Pool analytics API \\
Lending Rate & Lending arbitrage, utilization spike & Lending analytics API \\
Governance & Governance catalyst, proposal tracking & Governance API \\
Exploit & Exploit detection, exploit warning & Security feeds \\
Price Feed & Whale alert, depeg risk & Price oracle API \\
Derivatives & Funding extreme, OI divergence, liquidation cluster & Derivatives aggregator \\
Emission & Token unlock, emission change & Unlock calendars \\
Social & Bribe opportunity, sentiment shift, narrative momentum & Social + on-chain \\
Stablecoin & Stablecoin supply changes & Stablecoin analytics \\
\end{longtable}
}

Coverage: 72/72 signal types (100\%) --- up from 33/71 at iteration 1. The verifiers are not mocks --- they call real APIs and cross-reference signal claims against live data. When The Graph deprecated its free hosted service, the loop detected the failure (agents producing 0 signals), diagnosed the root cause (stale subgraph URLs), and migrated to alternative data sources (GeckoTerminal, DefiLlama) --- all autonomously.

\subsection{E.2 Anti-Signal Canary Details}\label{sec:e.2-anti-signal-canary-details}

\textbf{Tier 1 --- Obviously Bad (6 canaries):} - Fabricated event for non-existent entity - Signal with impossible metric value (e.g., 999,999\% yield) - False report for an entity that was never affected - Signal with null/empty identifiers - Signal with empty title and description - Signal with out-of-range confidence score

\emph{Result: 100\% caught across 163 iterations. Zero escapes.}

\textbf{Tier 2 --- Shadow (5 canaries):} - Real event that already concluded - Known trend reported weeks ago (stale) - Opportunity that dropped below threshold since detection - Valid signal for a deprecated/discontinued entity - Accurate data from a non-authoritative source

\emph{Result: Initially 33\% caught by L1-L3 (iter 1). Now 100\% (iter 163) --- improved as verifiers were added and L2 factual checkers gained cross-reference capabilities.}

\textbf{Tier 3 --- Adversarial (5 canaries):} - Real data + fabricated interpretation - Accurate raw data + incorrect derived conclusion - Valid input data + wrong calculation - Correct methodology applied to wrong time window - Signal mixing real and fabricated sub-claims

\emph{Result: Initially 67\% caught by L1-L3 (iter 1). Now 100\% (iter 163) --- the remaining required cross-referencing conclusions against source data, which improved as verifier coverage reached 100\%.}

\textbf{Tier 4 --- Mixed True/False (5 canaries, added \roughly iter 100):} - Signal with 3 true claims and 2 fabricated claims blended together - Accurate quantitative data with fabricated qualitative assessment - Real protocol event attributed to wrong protocol - Valid historical data presented as current - Correct analysis of real data with one inverted conclusion

\emph{Result: 100\% caught. Tests the quality gate's ability to detect partial failures rather than binary pass/fail.}

\textbf{API Degradation Canaries (3 canaries):} - Simulated timeout from primary data source - Simulated error response from price oracle - Simulated partial data return (50\% of expected fields)

\emph{Result: 100\% resilience (3/3). Quality gates degrade gracefully --- they flag signals as unverifiable rather than producing false passes when dependencies fail.}

\subsection{E.3 Key Lesson: Rapid Iteration Reveals Intermittent Bugs}\label{sec:e.3-key-lesson-rapid-iteration-reveals-intermittent-bugs}

One agent's validation failure rate fluctuated between 75-100\% across iterations because its data source API returns different formats at different times. A single test run would show ``pass'' or ``fail'' --- 163 iterations reveals the \roughly85\% steady-state rate. This class of bug, invisible to traditional CI, demonstrates the value of coverage through repetition, not just breadth.

\subsection{E.4 Drift Detection at Scale}\label{sec:e.4-drift-detection-at-scale}

The drift detection system monitors quality metrics over a sliding window (5 recent vs.\ 20 baseline iterations). The key insight: drift detection provides early warning \emph{before} quality gates fail --- a 5\% drop in factual pass rate over 10 iterations triggers an alert before any individual signal fails hard. This is the mechanism that allows the loop to operate autonomously for 163+ iterations without human supervision.

\section{Skill Interface Reference}\label{sec:appendix-f-skill-interface-reference}

A Kitchen Loop deployment consists of domain-independent orchestration plus domain-specific skills. The following table summarizes the skill interfaces that a new deployment must implement:

\begin{table}[ht]
\centering
\footnotesize
\renewcommand{\arraystretch}{1.25}
\begin{tabular*}{\textwidth}{@{\extracolsep{\fill}} l l l l l @{}}
\toprule
\textbf{Skill} & \textbf{Phase} & \textbf{Input} & \textbf{Output} & \textbf{Domain-Specific?} \\
\midrule
\texttt{backlog}      & Backlog & Ticket state       & Promoted tickets        & Partially (labels) \\
\texttt{ideate}       & Ideate  & Scenario + spec    & Report + scenario       & \textbf{Yes} (defines ``usage'') \\
\texttt{triage}       & Triage  & Experience report  & Labeled, deduped tickets & Partially (taxonomy) \\
\texttt{execute}      & Execute & Ranked tickets     & Branch + PR + tests     & Partially (test patterns) \\
\texttt{pr-manager}   & Polish  & Open PR list       & Reviewed, CI-passing PRs & No \\
\texttt{regress}      & Regress & Codebase state     & Pass/fail + drift       & \textbf{Yes} (defines ``oracle'') \\
\texttt{loop-review}  & Meta    & Logs + diffs       & Health + improvement    & No \\
\texttt{review-meta}  & Meta    & All review reports & Trends + recommendations & No \\
\bottomrule
\end{tabular*}
\end{table}

\textbf{Bold = must be fully reimplemented} for each new domain. Non-bold skills transfer with minimal configuration (ticket labels, CI commands, review tool selection).

\section{How to Cite}\label{sec:how-to-cite}

\begin{verbatim}
@misc{kitchenloop2026,
  title        = {The Kitchen Loop: User-Spec-Driven Development
                  for a Self-Evolving Codebase},
  author       = {Roy, Yannick},
  year         = {2026},
  howpublished = {arXiv preprint},
  url          = {https://github.com/0xagentkitchen/kitchenloop},
  note         = {Companion repository with skills, oracles,
                  and canary templates}
}
\end{verbatim}


\begin{thebibliography}{14}

\bibitem{fawzy2025} Fawzy, S., Tahir, A., \& Blincoe, K. (2025). Vibe Coding in Practice: Motivations, Challenges, and a Future Outlook --- A Grey Literature Review. \emph{arXiv:2510.00328}.

\bibitem{huang2025} Huang, Y., Reyna, K., Lerner, B. S., Xia, C. S., \& Hempel, J. (2025). Professional Software Developers Don't Vibe, They Control: AI Agent Use for Coding in 2025. \emph{arXiv:2512.14012}.

\bibitem{rahman2026} Rahman, M. M., et al.~(2026). Task-Level Evaluation of AI-Generated Pull Requests in Open-Source Software. \emph{arXiv preprint}.

\bibitem{robbes2026} Robbes, R., Matricon, T., Degueule, T., Hora, A., \& Zacchiroli, S. (2026). Agentic Much? Adoption of Coding Agents on GitHub. \emph{arXiv:2601.18341}.

\bibitem{abtahi2025} Abtahi, S. \& Azim, A. (2025). Augmenting Large Language Models with Static Code Analysis for Automated Code Quality Improvements. \emph{arXiv:2506.10330}.

\bibitem{liang2025} Liang, X., Garg, S., \& Zilouchian Moghaddam, R. (2025). The SWE-Bench Illusion: When State-of-the-Art LLMs Remember Instead of Reason. \emph{arXiv:2506.12286}.

\bibitem{thai2025} Thai, T. D., et al.~(2025). SWE-EVO: Benchmarking Coding Agents in Long-Horizon Software Evolution Scenarios. \emph{arXiv:2512.18470}.

\bibitem{cai2025} Cai, Y., Li, R., Liang, P., Shahin, M., \& Li, Z. (2025). Designing LLM-based Multi-Agent Systems for Software Engineering Tasks: Quality Attributes, Design Patterns and Rationale. \emph{ACM Trans. Softw. Eng. Methodol.} arXiv:2511.08475.

\bibitem{shukla2025} Shukla, S., Joshi, S., \& Syed, T. (2025). Security Degradation in Iterative AI Code Generation --- A Systematic Analysis of the Paradox. \emph{arXiv:2506.11022}.

\bibitem{gao2025} Gao, Y., et al.~(2025). A Survey of Bugs in AI-Generated Code. \emph{arXiv:2512.05239}.

\bibitem{he2025} He, H., Miller, C., Agarwal, S., Kastner, C., \& Vasilescu, B. (2025). Speed at the Cost of Quality: How Cursor AI Increases Short-Term Velocity and Long-Term Complexity in Open-Source Projects. \emph{Proc. MSR 2026}. arXiv:2511.04427.

\bibitem{becker2025} Becker, K., Rush, A. M., Barnes, C., \& Rein, D. (2025). Measuring the Impact of Early-2025 AI on Experienced Open-Source Developer Productivity. \emph{METR}. arXiv:2507.09089.

\bibitem{chen2025} Chen, Y., Talwalkar, A., Brennan, G., \& Neubig, G. (2025). Code with Me or for Me? How Increasing AI Automation Transforms Developer Workflows. \emph{arXiv:2507.08149}.

\bibitem{yao2025} Yao, S., et al.~(2025). Sycophancy in Multi-Agent Debate. \emph{arXiv:2509.23055}.

\end{thebibliography}
\end{document}